\documentclass[aps,prb,showpacs,
floatfix,
amsmath,amssymb,superscriptaddress,reprint]{revtex4-1}
\usepackage{bbold}
\usepackage{color}
\usepackage{natbib}
\usepackage{epsfig}
\usepackage{setspace}
\usepackage{amsmath}
\usepackage{amssymb}
\usepackage{verbatim}
\usepackage{tikz}
\usepackage{bibentry}
\usepackage{physics}
\usepackage{hyperref}
\usepackage{cancel}
\usepackage{subcaption}
\usepackage{graphicx}
\usepackage{array}
\usepackage[font=small,labelfont=bf,
   justification=justified,
   format=plain]{caption}
\def\etal{{\em{et al. }}}

\def\bk{{\mathbf{k}}}

\definecolor{scarred}{rgb}{0.75,0.0,0.0}
\begin{document}
\usetikzlibrary{arrows.meta,decorations.markings}
\preprint{APS/123-QED}
\title{Kondo effect in a non-Hermitian, $\mathcal{PT}$-symmetric Anderson model with Rashba spin-orbit coupling}
\author{Vinayak M Kulkarni}
\affiliation{Theoretical Sciences Unit, Jawaharlal Nehru Centre for Advanced Scientific Research Jakkur, Bangalore - 560064,India}
\author{Amit Gupta}
\affiliation{ Department of Physics, M. R. M. College, Lalit Narayan Mithila University, Darbhanga, Bihar 846004, India}
\author{N. S. Vidhyadhiraja}
\affiliation{Theoretical Sciences Unit, Jawaharlal Nehru Centre for Advanced Scientific Research Jakkur, Bangalore - 560064,India}

\begin{abstract}
The non-interacting and non-Hermitian, parity-time ($\mathcal{PT}$)-symmetric Anderson model exhibits an exceptional point (EP) at a non-Hermitian coupling $g=1$, which remains unrenormalized in the presence of interactions (Lourenco et al, Phys.\ Rev.\ B {\bf 98}, 085126 (2018)), where the EP was shown to coincide with the quantum critical point (QCP) for Kondo destruction. In this work, we consider a quantum dot hybridizing with metallic leads having Rashba spin-orbit coupling ($\lambda$). 
We show that for a non-Hermitian hybridization, $\lambda$ can renormalize the exceptional point even in the non-interacting case, stabilizing $\mathcal{PT}$-symmetry beyond $g=1$. Through exact diagonalization of a zero-bandwidth, three-site model, we show that the quantum critical point and the exceptional point bifurcate, with the critical point for Kondo destruction at $g_c=1$, and the exceptional coupling being $g_{\scriptscriptstyle{EP}} > 1$ for all $U\neq 0$ and $\lambda\geq 0; \lambda\neq U/2$. On the line $\lambda=U/2$, the critical point and the EP again coincide at $g_c=g_{\scriptscriptstyle{EP}}=1$. The full model with finite bandwidth leads is investigated through the slave-boson approach, using which we show that, in the strong coupling regime, $\lambda$ and interactions co-operate in strongly reducing the critical point associated with Kondo destruction, below the $\lambda=0$ value.
\end{abstract}

\maketitle
\section{Introduction}

    Conventional quantum theory postulates that every physical observable may be represented by a Hermitian operator, since Hermiticity ensures that the eigenvalues of the corresponding operators would be real. 
A generalization to the existing quantum postulates can be made with consideration of $\mathcal{PT}$-symmetric orthonormal set of eigenstates{\cite{zhao2015robust,rosas2018bi}}. These states preserve the norm and form a complete set which also allows for an unambiguous definition of expectation of physical observables. Physically, non-Hermitian models represent open quantum systems\cite{rotter2009non,echeverri2019comparative,wang2021application}. These models can exhibit eigenvalue degeneracies at certain values of non-Hermitian parameters, called exceptional points (EPs), which correspond to quantum phase transitions of the level-crossing type. Concomitantly, these points also show a break down of $\mathcal{PT}$-symmetry. Beyond such exceptional points, the eigenvalues develop finite imaginary parts, and the norm of the corresponding eigenvector is not conserved.  Hence, probabilities of eigenstates oscillate with a decay factor as a function of non-Hermitian strength. The emergence of imaginary eigenvalues may be associated with a loss of bound states and a crossover to scattering states of open quantum systems\cite{bender2007making,matzkin2006non,jones2007scattering}. 
    
    Many theoretical studies of non-Hermitian 
Hamiltonians have been motivated by experiments that realize such models in open systems with balanced gain and loss \cite{schi,ruter2010observation,krinner2017two,rotter2009non,longhi2018parity}. Recent experiments in cold atoms have observed level crossing like transitions even though many-body interactions are present in the system.
The observations of gain and loss due to depletion of atoms have been attributed to the phenomenon of continuous quantum Zeno effect \cite{cold1,cold2,cold3,cold4}. 
Conventional Kondo type systems such as impurities coupled to baths have been realised in controlled environments like 
atoms in harmonic traps\cite{localized,kanasz2018exploring,zhang2020controlling,nishida2016transport,nakagawa2015laser,bauer2013realizing}, where the small number of excited states mimic magnetic impurities and the atoms in the ground state provide the bath. 
In  experiments where Rashba type spin-obit coupling is generated synthetically, induced artificial magnetic fields break parity and
time reversal symmetries. It is important to note that when both of these symmetries break, SU(2) symmetry will also be violated. Exceptional points may also lead to exceptionally sensitive sensors ~\cite{miller2017exceptional,wiersig2020review}  since for an EP of order $n$, any perturbation of strength $\epsilon$ leads to a splitting of the levels ($\Delta_L$) that is proportional to the n$^{th}$ root of $\epsilon$, which is in contrast to that of a diabolic point where $\Delta_L\sim\epsilon$.

    
  Open quantum systems are generally investigated through master equation techniques with various kinds of noise terms as discussed by Plenio and Knight~\cite{jump}.  
   Quantum criticality in $\mathcal{PT}$-symmetric
effective field theories have been studied in the sine-Gordon model\cite{ashida2017parity} which has been shown to describe the transition between a Mott-insulator to  a Tomonaga-Luttinger liquid \cite{criti}.
 The $\mathcal{PT}$-symmetry breaking transition has  also been studied in a classical system, where critical behavior, similar to that found in quantum systems, has been observed through numerical calculations \cite{bender2007spontaneous}. Classical spin chains with imaginary fields also exhibit similar phase transitions\cite{castro2009spin}. Another work in interacting quantum many-body systems discusses a novel way of doing path integrals for such systems to capture the Anderson localisation transition with non-Hermitian disorder \cite{hatano1996localization}. 
 
Lourenco et al\cite{loure} considered a real-space parity-time symmetric  model comprising a correlated impurity connected to left and right leads through a non-hermitian hybridization coupling. They have employed perturbative renormalization group (RG) approach to investigate the exceptional points in the strong coupling regime. The non-interacting exceptional point was shown to coincide with the critical point for Kondo destruction and also found to remain invariant under RG flow. 
Nakagawa \etal{\cite{poor}} have considered a non-Hermitian (NH) Kondo model that is not $\mathcal{PT}$-symmetric. It is an extension of the standard Kondo model to one with complex spin exchange couplings and the justification for these non-Hermitian terms is given through Lindbladian dynamics. Standard two-loop poorman RG yields a phase transition, as seen through RG reversion, which occurs at a very small value of the complex coupling, since there is no symmetry and there is a 
local-moment type fixed point. These results have been further supported through Bethe ansatz calculations.

Our interest is to explore exceptional and quantum critical points in a non-hermitian quantum many-body system subjected to a decohering term such as the Rashba spin-orbit coupling (RSOC). The interplay of RSOC and interactions has been investigated extensively in the Hermitian case  using RG methods such as poorman scaling and numerical renormalization group\cite{malecki,zarea,sandler2,sandler3,Zitko}.  Other special, but Hermitian, cases such as an impurity in graphene, nanoribbons, and an impurity in a topological insulator have been considered \cite{Zitko,hu2011magnetic} and
 through a mapping onto the pseudogap Anderson model, a quantum phase transition has been shown to occur. The SO coupling, particularly the Rashba type, breaks parity symmetry in conventional models represented by an angular momentum basis. Concomitantly, this also leads to the generation of  $\mathcal{PT}$ symmetric channels and hence a Dzyaloshinskii-Moriya (DM) interaction in the effective model of impurity subspace, which leads us to consider interesting possibilities. If we consider NH coupling to these $\mathcal{PT}$ channels, there is an emergent parity-time breaking DM interaction which may modify the exceptional points  of the NH model. In the conventional Hermitian case, closed systems may exhibit phase transitions as a function of SO interactions only for some special lattices like a honeycomb lattice as seen in e.g.\ Bi$_{2}$Se$_{3}$ and Bi$_{2}$Te$_{3}$. However, in NH systems, one can expect phase transitions driven by the imaginary interaction generically irrespective of the lattice.
With the prospect of the above possibilities, we ask the following questions: (a) Can RSOC renormalize the exceptional point in the non-interacting case? (b) What is the combined effect of interactions and RSOC on Kondo destruction and the exceptional points? For investigating these, we have considered a single level quantum dot connected to a bath which has RSOC  (which is important for realizing $\mathcal{PT}$ symmetry\cite{nejati2017kondo,localized}). The model is also motivated by recent experiments\cite{localized} where singlet and triplet scales in open conditions have been measured.

We set up the full, non-Hermitian, single impurity Anderson model, and establish ${\mathcal{PT}}-$symmetry, first in a simplified zero-bandwidth, three-site model, and subsequently in the full model. Using exact diagonalization, we show the emergence of distinct quantum critical and exceptional points in the three-site model. Subsequently, the full model in the non-interacting case is solved through exact diagonalization, Green's functions methods, and total energy calculations. In order to understand the effect of interactions, we utilise the slave-boson method, and show the co-operative interplay of non-Hermitian coupling and RSOC ($\lambda$) in inducing Kondo destruction. 
We show that a finite $\lambda$ protects ${\mathcal{PT}}-$symmetry by pushing the exceptional point beyond the $\lambda=0$ value. In the strong coupling regime, a quantum phase transition occurs between a Kondo screened phase and an unscreened moment at a critical non-Hermitian coupling $g_c = 1$ for $\lambda=0$. With increasing $\lambda$, the critical coupling decreases monotonically showing a strong renormalization of the QCP due to RSOC.
    
The paper is organized as follows: The following section introduces the model and formalism. Section III introduces a simplified three-site, zero-bandwidth model using which ${\mathcal{PT}}$-symmetry is analyzed and 
exceptional points are found in closed form. Through exact diagonalization of the interacting three-site model in Fock space, we show the bifurcation of the exceptional point and the quantum critical point. We present the results and discussion for the full, finite bandwidth leads, model in section IV, and conclude in the final section with a short discussion and open questions.

\section{Model and formalism}

As mentioned in the introduction, we have chosen to work with a single impurity Anderson model (SIAM) comprising a single non-degenerate level quantum dot system connected to an electron reservoir,
for which the Hamiltonian, $H$, is given in standard notation as
\begin{equation}
    H_{SIAM} = H_{0} + H_{\rm d} + H_{\rm hyb}\,,
\end{equation}
    where, the two-dimensional conduction band reservoir may be represented by $H_0 = \sum_{\bk\sigma}\epsilon_\bk c^\dagger_{\bk\sigma} c^{\phantom{\dagger}}_{\bk\sigma}$ and the isolated quantum dot is given  by
$H_{\rm d} = \sum_{\sigma}\epsilon_{d} d^{\dagger}_{\sigma}d^{\phantom{\dagger}}_{\sigma}+Un_{d\uparrow}n_{d\downarrow}$. The 
hybridization term is given by $H_{hyb}=\sum_{\bk\sigma}V_{\bk\sigma}(c^{\dagger}_{\bk\sigma}d^{\phantom{\dagger}}_{\sigma}+{\rm h.c.})$.

 The presence of spin-orbit coupling in a two-dimensional conduction electron bath  has been 
considered previously by several groups \cite{Zitko,zarea,sandler2,sandler3}. In this work, we have investigated the interplay of Rashba type spin-orbit coupling (SOC) with the presence of non-Hermitian, but PT-symmetric terms in the Hamiltonian. Defining $\psi_\bk = \left( c_{\bk\uparrow}\;c_{\bk\downarrow}\right)^T$, the SOC term is given by: 
\begin{equation}
    \begin{split}
        H_{\rm RSO} & = \lambda\sum_{\bk} \psi^{\dagger}_{\bk}(\bk\times \vec{\sigma})_z\psi^{\phantom{\dagger}}_{\bk}\\
        & =\lambda \sum_{\bk}k \left( e^{i\theta_{\bk}}c^{\dagger}_{\bk\uparrow}c_{\bk\downarrow}^{\phantom{\dagger}}+{\rm h.c.}\right)
    \end{split}
\end{equation}
where $k=|\bk|$ and $\theta_\bk= \tan^{-1}(-k_x/(-k_y))$.
 The action of the parity operator is:
\begin{equation}
    \begin{split}
        \mathcal{P}: \sigma_{x}\psi_{k}  \implies c_{k\uparrow}\rightarrow c_{k\downarrow}\,,
    \end{split}
\end{equation}
and using the above, we see that $H_{\rm RSO}$ is not invariant under parity transformation.

The conduction band terms, namely $H_0$ and $H_{\rm RSO}$ may be combined\cite{zarea}, which leads to the emergence of chiral conduction bands. This is accomplished using an angular momentum expansion for the conduction band operators, followed by a unitary transformation as:
\begin{equation}
\label{eq:AMtrans}
    c_{\bk\sigma} = c_{{\scriptscriptstyle{k_x k_y}}\sigma} = \frac{1}{\sqrt{2\pi k}}\sum_{m=-\infty}^{\infty}
    c_{km\sigma}\exp(-i m \theta_\bk)\,,
\end{equation}
where $k=|\bk|$. The inverse transform is defined as $c_{km\sigma}=\sqrt{\frac{k}{2\pi}}\int^{2\pi}_{0} d\theta_{\bk} c_{\bk \sigma} e^{im\theta_{\bk}}$.
Substituting the above expansion\cite{zarea} (equation~\ref{eq:AMtrans}) into the Hamiltonian, and assuming an isotropic dispersion, such as $\epsilon_\bk = \hbar^2 k^2/2m$, the $H_0$ becomes:
\begin{equation}
  \label{eq:H0m}
        H_{0}=\sum_{\bk\sigma}\epsilon_{\bk}c^{\dagger}_{\bk\sigma}c^{\phantom{\dagger}}_{\bk\sigma} 
        =\sum_{km\sigma}\tilde{\epsilon}_{k}c^{\dagger}_{km\sigma}c^{\phantom{\dagger}}_{km\sigma}
\end{equation}
where $\tilde{\epsilon}_{k}=\epsilon_{\bk}/k$. Further, with the same transformation, the RSO term transforms to:
\begin{equation}
\label{eq:hsrom}
    H_{\rm RSO}=\lambda\sum_{km} \left(c^\dag_{k,m\uparrow} 
    c^{\phantom{\dag}}_{k\,m+1\downarrow} + {\rm h.c.}\right)
\end{equation}
In Appendix-A, we develop a non-Hermitian Anderson model Hamiltonian by combining the above terms and a few extra terms invoking Lindladian dynamics, and the resulting Hamiltonian has the following form:
\begin{equation}
   \begin{split}
    H &=\sum_{kh\eta}\tilde{\epsilon}_{kh}^{\phantom{\dagger}} c_{kh\eta}^\dagger c_{kh\eta}^{\phantom{\dagger}}  
  +\sum_{k\eta h} X_{k\eta h}\left( c^{\dagger}_{k\eta h}d_h^{\phantom{\dagger}}+ {\rm h.c.}\right)\\
 &+\epsilon_d \sum_{h} n_{dh}
    + Un_{d+}n_{d-}
    \end{split}
    \label{eq:Canderson}
\end{equation}
where $h$ is a 'chiral' quantum number and can be thought of as a pseudospin index, $\eta\in\{L,R\}$ is the channel index, which is, in fact the $j_m=m+\sigma$ index in the angular momentum representation.  The hybridization coefficients are $X_{kLh} = \sqrt{2}|X_k|e^{i\phi_k}$ and $X_{kRh} = -hX_{kLh}^*$.
The model obtained in the rotated basis above has the interpretation of a dot connected to two leads, and appears very similar to the {\em real-space} model considered by 
Lourenco et al~\cite{loure}. The main difference is that the real-space model had non-Hermitian coupling to just two sites that were directly connected to the dot. In our case, the hybridization elements, $X_{k\eta}$, being complex, render the Hamiltonian non-hermitian when $\phi_k\ne 0,\pi$ and for all $k$.

Before we investigate the full model in equation~\ref{eq:Canderson}, we
have considered a three-site, zero-bandwidth model that has a very similar structure
as equation~\ref{eq:Canderson}. We will see that the symmetry class, exceptional points etc can be
obtained easily and exactly, and hence is very instructive. Furthermore, a generalization of the symmetry analysis to the full model will be straightforward.

\section{Three-site, zero bandwidth model in the chiral basis}
  Consider a simplification of the above model (equation~\ref{eq:Canderson}), where the leads are replaced by single sites with two orbitals each.
 \begin{equation}
 \begin{split}
    H &=\sum_{h\eta}\epsilon_{h}^{\phantom{\dagger}} c_{h\eta}^\dagger c_{h\eta}^{\phantom{\dagger}}  
  +\sum_{\eta h} X_{\eta h}\left( c^{\dagger}_{\eta h}d_h^{\phantom{\dagger}}+ {\rm h.c.}\right) \\
  & +\epsilon_d \sum_{h} n_{dh} + Un_{d+}n_{d-}
   \end{split}
\label{eq:Canderson_app}
\end{equation}
where $\epsilon_h = \epsilon + h\lambda$, $X_{Lh} = V e^{i\phi}$, $X_{Rh} = -hX_{Lh}^*$ and $V$ is a real number ($V\in {\mathcal{R}}$). The exceptional points may be found in terms of $\phi$ or a coupling $g$ defined as the ratio of the imaginary part to the real part of the hybridization, which is simply $g=\tan\phi$. 
  Using $\psi=\begin{pmatrix}c_{L+}&c_{R+}&d_{+}&d_{-}&c_{L-}&c_{R-}\end{pmatrix}^T$, we can write the above Hamiltonian as 
 \begin{equation}
     H=\psi^\dagger {\mathcal{H}} \psi + Un_{d+}n_{d-}
 \end{equation}
 where 
 \begin{equation}
   {\mathcal{H}} =
   \begin{pmatrix}
   \epsilon_+ & 0 & X_{L+} & 0 & 0 & 0 \\
   0 & \epsilon_+ & X_{R+} & 0 & 0 & 0 \\
   X_{L+} & X_{R+} & \epsilon_d & 0 & 0 & 0 \\
   0 & 0 & 0 & \epsilon_d & X_{L-} & X_{R-} \\
   0 & 0 & 0 & X_{L-} & \epsilon_- & 0 \\
   0 & 0 & 0 & X_{R-} & 0 & \epsilon_-
   \end{pmatrix}
 \end{equation}
 This matrix is block-diagonal, since the chiral channels do not mix in the absence of interaction, i.e for $U=0$. It is also non-Hermitian, but symmetric, i.e ${\mathcal{H}}^\dagger \neq {\mathcal{H}}$, but ${\mathcal{H}}^T={\mathcal{H}}$. Now, we explore the non-interacting case first, before moving on to $U\neq 0$.
 
 \subsection{Non-interacting case: \texorpdfstring{$U=0$}{Lg}}

 Keeping $X_{\eta h}$ general, we want to find conditions so that the eigenvalues are real. The eigenvalues of the above Hamiltonian ($\Lambda$) (for the special case of $\epsilon=\epsilon_d=0$ and $U=0$), are given by: 
\begin{equation}
    \begin{split}
        &\Lambda  = \pm \lambda \\
        &\Lambda^2 -(\pm\lambda)\Lambda - (X_{L\pm}^2+
        X_{R\pm}^2) = 0
    \end{split}
\end{equation}
Thus the condition that determines real eigenvalues is
\begin{equation}
    \cos2\phi \ge -\frac{\lambda^2}{8V^2}
    \label{eq:nint_real}
\end{equation}
which reduces to $\phi \le \pi/4$ in the absence of RSOC, while if $\lambda\neq 0$, the condition is
as given above, so RSOC stabilizes $\mathcal{PT}$-symmetry by increasing the range of $\phi$ to beyond $\pi/4$.
And if $\lambda\ge 2\sqrt{2}V$, $\mathcal{PT}$-symmetry can not be broken for any $\phi$. Thus, for $\lambda\le 2\sqrt{2}V$, the exceptional point is given by,
\begin{equation}
    \phi_{\scriptscriptstyle{EP}} = \pi/4 + \frac{1}{2}\sin^{-1} \frac{\lambda^2}{8V^2}\,,
 \end{equation}
or equivalently in terms of $g$, the EP is given by
\begin{equation}
    g_{\scriptscriptstyle{EP}} = \tan\phi_{\scriptscriptstyle{EP}} = \sqrt{\frac{1+\lambda^2/8V^2}{1-\lambda^2/8V^2}}\,.
\end{equation}
So, the minimal condition necessary for 
real eigenvalues is $X_{Rh}^2+X_{Lh}^2 \in {\mathcal{R}}$ for $h=\pm$.
From an inspection of the Hamiltonian, a mapping that yields ${\mathcal{H}}\rightarrow {\mathcal{H}}^\dagger$ is
\begin{equation}
\psi=\begin{pmatrix}
c_{L+}\\
c_{R+}\\
d_{+}\\
d_{-}\\
c_{L-}\\
c_{R-}
\end{pmatrix}\rightarrow
\begin{pmatrix}
-c_{R+}\\
-c_{L+}\\
d_{+}\\
d_{-}\\
c_{R-}\\
c_{L-}
\end{pmatrix}
\end{equation}
 This implies that the matrix representation of the metric operator, $\eta$, that should yield this transformation should be
 \begin{equation}
 \eta = 
 \begin{pmatrix}
 0 & -1 & 0 & 0 & 0 & 0 \\
 -1 & 0 & 0 & 0 & 0 & 0 \\
 0 & 0 & 1 & 0 & 0 & 0\\
 0 & 0 & 0 & 1 & 0 & 0 \\
 0 & 0 & 0 & 0 & 0 & 1 \\
 0 & 0 & 0 & 0 & 1 & 0
 \end{pmatrix}    
 \end{equation}
 such that
 $\eta^2=\mathbb{1}$, and thus $\eta^\dagger=\eta^{-1}=\eta$, which is unitary.
 Indeed, we find that
 \begin{equation}
     \eta {\mathcal{H}} \eta^{-1} = {\mathcal{H}}^\dagger\,
     \label{eq:pherm}
 \end{equation}
implying that $H$ is pseudohermitian.
Since ${\mathcal{H}}$ is also symmetric, i.e ${\mathcal{H}}^T={\mathcal{H}}$, the pseudohermiticity is identical to
 ${\mathcal{PT}}$-symmetry~\cite{PhysRevResearch.2.033022}.
 We can also make a statement about the left ($\Psi_L$) and right eigenvectors ($\Psi_R$), as follows.
 Since ${\mathcal{H}}\Psi_{\alpha R}=E_\alpha\Psi_{\alpha R}$, where $E_\alpha$ is the $\alpha^{\rm th}$ eigenvalue, and given the property
 \ref{eq:pherm}, we can see that
 \begin{equation}
      {\mathcal{H}}^\dagger (\eta \Psi_{\alpha R}) = E (\eta \Psi_{\alpha R})
 \end{equation}
 and hence the left eigenvector corresponding to the complex conjugate eigenvalue $E_\alpha^*$ would be $\Psi_{\alpha L}=\eta\Psi_{\alpha R}$.
We can also construct the metric operator in the second quantized form, as
\begin{equation}
    \begin{split}
    \hat{\eta} = &\psi^\dagger \eta^{\phantom{\dagger}} \psi^{\phantom{\dagger}} \\
      = & -\left(c^\dagger_{R+} c^{\phantom{\dagger}}_{L+} + {\rm h.c}\right)
      +\left(c^\dagger_{R-}c^{\phantom{\dagger}}_{L-} + {\rm h.c}\right)  \\
      & + \left(d_+^\dagger d_+^{\phantom{\dagger}} +
    d_-^\dagger d_-^{\phantom{\dagger}}\right)
    \end{split}
\end{equation}
For the Hamiltonian to be pseudoHermitian,
it is easy to see from equation~\ref{eq:pherm} that the condition to be satisfied is
\begin{equation}
    \left[H + H^\dagger,\hat{\eta}\right] = 0
\label{eq:oppherm}
\end{equation}
and indeed we see that this condition is satisfied as shown below.
\begin{equation}
    \begin{split}
        \left[H,\hat{\eta}\right] & = 
        \left(X_L+X_{R+}\right)\left((c_{L+}^\dagger + c_{R+}^\dagger)d_+^{\phantom{\dagger}}
        - {\rm h.c} \right) \\
        &+ \left(X_L-X_{R-}\right)\left((c_{L-}^\dagger - c_{R-}^\dagger)d_-^{\phantom{\dagger}}
        - {\rm h.c} \right)
    \end{split}
\end{equation}
With the above result, we see that the condition for pseudohermiticity (equation~\ref{eq:oppherm}) is satisfied since
\begin{equation}
    Re(2X_L+X_{R+}-X_{R_-}) = 2V -V -V=0\,.
\end{equation}
Thus, with the combination of pseudohermiticity and symmetric form, we establish ${\mathcal{PT}}$-symmetry.

\subsection{Interacting case: \texorpdfstring{$U>0$}{Lg}}
We perform exact diagonalization of the three-site model  for $U\neq 0$ (equation~\ref{eq:Canderson_app})  in the Fock space to get an insight into the combined effect of $U$, $\lambda$ and non-hermiticity on the phase diagram. The exceptional point is found as usual from the emergence of a non-zero imaginary part in the eigenvalues of the Hamiltonian, while a quantum phase transition would be signalled by a crossing of the real-valued ground state and the first excited state eigenvalues.

 \begin{figure}[b]
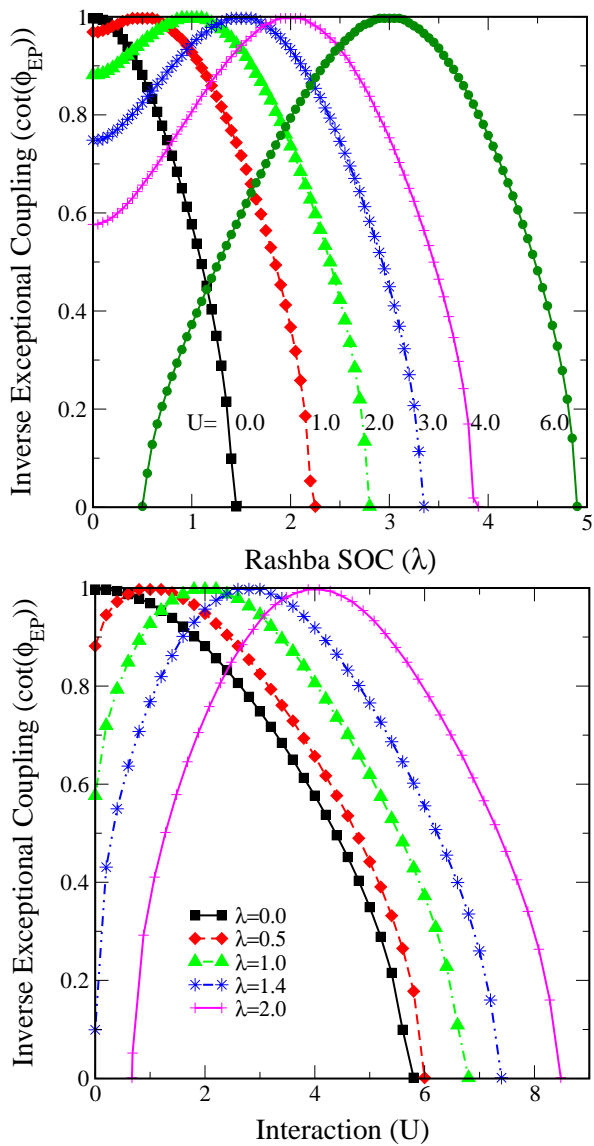

    \centering
\includegraphics[scale=0.55,trim=0 0 0 0, clip]{phi_EP_vs_SOC_var_U.eps}
\includegraphics[scale=0.55,trim=0 0 0 0, clip]{phi_EP_vs_U_var_SOC.eps}
\caption{The inverse exceptional coupling, $g_{\scriptscriptstyle{EP}}^{-1}=\cot(\phi_{\scriptscriptstyle{EP}})$ as a function of (top panel) $\lambda$ for various $U$ values and (bottom panel) $U$ for various $\lambda$ values. }
\label{fig:EP_vs_SOC}
\end{figure}

We have seen in the non-interacting case (refer equation~\ref{eq:nint_real}) that the spin-orbit coupling stabilizes
${\mathcal{PT}}$-symmetry, in the sense that $\phi_{\scriptscriptstyle{EP}}$ increases from $\pi/4$ at $\lambda=0$ to $\pi/2$ at $\lambda_c=2\sqrt{2}V$ beyond which the
exceptional point does not arise, implying that the ${\mathcal{PT}}$-symmetry does not break. As mentioned before, the
non-Hermitian coupling at the exceptional point defined as $g_{\scriptscriptstyle{EP}}=\tan(\phi_{\scriptscriptstyle{EP}})$ increases from $1$ to $\infty$. Now, we can investigate the exceptional points in the presence of $U$ and $\lambda$.

The top panel of figure~\ref{fig:EP_vs_SOC} shows the inverse exceptional coupling, i.e  $g^{-1}_{\scriptscriptstyle{EP}}=\cot(\phi_{\scriptscriptstyle{EP}})$ as a function of $\lambda$ for various $U$ values, while the bottom panel shows the same as a function of $U$ for various $\lambda$ values. We see that for $U=0$, the exceptional points shift to higher values ($\phi_{\scriptscriptstyle{EP}}>\pi/4$ or $g> 1$) upon increasing $\lambda$, which is consistent with that found in the non-interacting case. The effect of finite interactions is to enhance the critical $\lambda$ beyond which ${\mathcal{PT}}-$symmetry is violated. Eventually for large $U$, the ${\mathcal{PT}}-$symmetry is unbroken upto a critical $\lambda$. Interestingly, we notice that at $\lambda=U/2$, the exceptional point reverts back to $\phi_{\scriptscriptstyle{EP}}=\pi/4$, which is just the same as the non-interacting value, thus negating the effect of $\lambda$ and $U$ completely. The implication is that, for a fixed $U$, if $\lambda=U/2$, the exceptional point is the same as that in the non-interacting, and zero SOC case, and is hence unrenormalized.
The bottom panel shows a similar behaviour as the top panel, with the roles of $U$ and $\lambda$ reversed, and again the $\lambda=U/2$ points are seen to be special unrenormalized points. 

\begin{figure}[h]
    \centering
\includegraphics[scale=0.55,trim=0 0 0 0, clip]{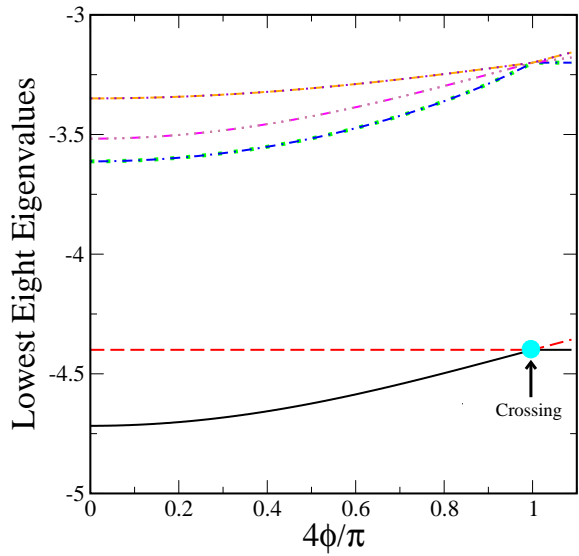}
\caption{The lowest eight eigenvalues as a function of the non-hermiticity parameter, $\phi$, for $U=4.0$, $\lambda=1.2$, and $\epsilon_d=-U/2$. The cyan dot shows the first crossing of the ground state and the first excited state, and is thus identified as the quantum critical point. }
\label{fig:evcross}
\end{figure}
In figure~\ref{fig:evcross}, we have shown the eight lowest eigenvalues as a function of $\phi$ for $U=4.0$, $\epsilon_d=-U/2$, $\lambda=1.2$, in the
${\mathcal{PT}}-$symmetry unbroken regime. The ground state, being adiabatically continuous with the hermitian case ($\phi=0$) is identified as the Kondo screened phase. This is also confirmed by examining the ground state eigenvector in the Fock space. The first crossing of the ground state and the first excited state eigenvalue, identified as the quantum critical point of the many-body level crossing type, remains at $\phi=\pi/4$ or $g=1$ for all $U\neq 0$ and any $\lambda$. 
We have confirmed through an eigenstate analysis also that the QCP represents a transition from a Kondo screened singlet phase to a local moment phase. 

\begin{figure}[htbp]
    \centering
\includegraphics[scale=0.60,trim=0 0 0 0, clip]{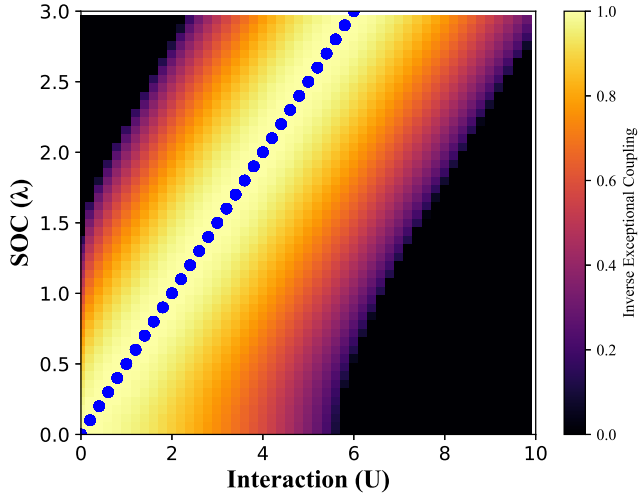}
\caption{The exceptional point phase diagram for the three site model in the interacting regime is shown. The colorbar shows the inverse exceptional coupling. The solid circles represent the line $\lambda=U/2$, on which the exceptional point remains at $\phi=\pi/4$ or $g=1$. Everywhere else the exceptional point is strongly renormalized by interactions and SOC. In the dark regions (low $U$, high $\lambda$ and high $U$, low $\lambda$), the inverse exceptional coupling vanishes, implying
$g_{\scriptscriptstyle{EP}}\rightarrow \infty$, and hence ${\mathcal{PT}}$-symmetry is never violated.}
\label{fig:phase_finite_U}
\end{figure}

The exceptional point phase diagram in the $U-\lambda$ plane is shown in figure~\ref{fig:phase_finite_U}. The colour represents inverse exceptional coupling, $g^{-1}_{\scriptscriptstyle{EP}}$. The extent of renormalization of the exceptional point due to interactions and SOC is indicated by the darkness of the colour. As mentioned above, the solid circles represent the line $\lambda=U/2$, where the exceptional point is completely unrenormalized with respect to $\lambda=U=0$. 
Finally, in the $U-\lambda$ plane, the Kondo destruction critical point for all $U\neq 0$ and any $\lambda$ coincides with the exceptional point for $\lambda=U/2$, while the exceptional point gets strongly renormalized away from this line, and does not even exist for $U/\lambda\gg 1$ and $U/\lambda\ll 1$.

\section{Full model: Results and discussion}
Taking cues from the solution of the three-site model, we can now explore the eigenvalues, and the symmetry of the full model, i.e equation~\ref{eq:Canderson}. Again, using
$\psi_k=\begin{pmatrix}c_{kL+}&c_{kR+}&d_{+}&d_{-}&c_{kL-}&c_{kR-}\end{pmatrix}^T$, we can write the Hamiltonian of equation~\ref{eq:Canderson} as 
 \begin{equation}
     H=\sum_k \psi_k^\dagger {\mathcal{H}}_k \psi_k + \epsilon_d\sum_h n_{dh}+ Un_{d+}n_{d-}
 \end{equation}
 where 
 \begin{equation}
   {\mathcal{H}}_k =
   \begin{pmatrix}
   \epsilon_{kL+} & 0 & X_{kL+} & 0 & 0 & 0 \\
   0 & \epsilon_{kR+} & X_{kR+} & 0 & 0 & 0\\
   X_{kL+} & X_{kR+} & 0 & 0 & 0 & 0 \\
   0 & 0 & 0 &  0 & X_{kL-} & X_{kR-} \\
   0 & 0 & 0 & X_{kL-} & \epsilon_{kL-} & 0 \\
   0 & 0 & 0 & X_{kR-} & 0 & \epsilon_{kR-}
   \end{pmatrix}
 \end{equation}
We observe from the form of the matrix above that the metric operator can be generalized from the case of the three-site model as 
\begin{equation}
    \begin{split}
    \hat{\eta} =& \sum_k \psi_k^\dagger \eta^{\phantom{\dagger}} \psi^{\phantom{\dagger}}_k \\
      = & \sum_k\left[-\left(c^\dagger_{kR+} c^{\phantom{\dagger}}_{kL+} + {\rm h.c}\right)
      +\left(c^\dagger_{kR-}c^{\phantom{\dagger}}_{kL-} + {\rm h.c}\right)  \right]\\
      & + \left(d_+^\dagger d_+^{\phantom{\dagger}} +
    d_-^\dagger d_-^{\phantom{\dagger}}\right)
    \end{split}
\end{equation}
which will again yield
\begin{equation}
    \left[H + H^\dagger,\hat{\eta}\right] = 0\,,
\label{eq:fullpherm}
\end{equation}
thus showing that the Hamiltonian is pseudohermitian even for $U\neq 0$, and since
$H=H^T$, we can identify the symmetry as being equivalent to $\mathcal{PT}$ symmetry. 

\subsection{Exact Diagonalization of the non-interacting model}

An exact diagonalization of the above model (for $U=0$ and $\epsilon_{kLh} = \epsilon_{kRh} $ and the hybridization elements assumed to be $k$-independent) yields the following equation for eigenvalues:
\begin{equation}
    (\epsilon_d - \Lambda) - \sum_k \frac{(X_{Lh}^2 + X_{Rh}^2)}{\epsilon_{kh} - \Lambda} = 0 \;\;\;\;{\rm for}\;h=\pm\,.
    \label{eq:nieig}
\end{equation}
Thus the possibility of  real eigenvalues exists if 
$(X_{Lh}^2 + X_{Rh}^2)\in {\mathcal{R}}$. 

 \begin{figure}[h]
    \centering
\includegraphics[scale=0.45,trim=0 0 0 0, clip]{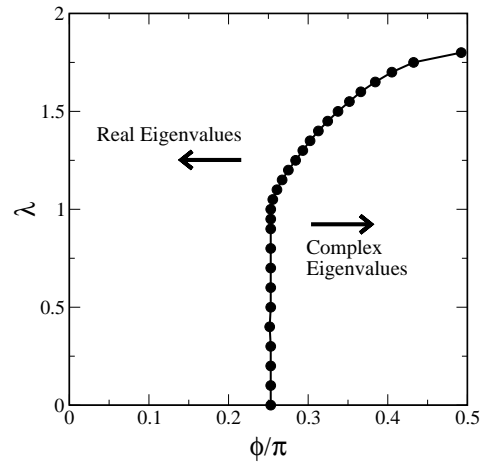}
     \caption{Phase diagram in the $\phi-\lambda$ plane is shown for the non-interacting case ($U=0$). The solid circles represent exceptional points, where certain real eigenvalues coalesce and complex conjugate eigenvalues emerge.  }
      \label{fig:phase_nint}
\end{figure}
The phase diagram for $U=0$ in the $\phi-\lambda$ plane is shown in figure~\ref{fig:phase_nint}.
The solid curve is the line of exceptional points, which shows that increasing spin-orbit coupling enhances the range of $\phi$ within which real eigenvalues are obtained, and hence SOC stabilizes ${\mathcal{PT}}$ symmetry. The exceptional point for $\lambda=0$ is at $\phi=\pi/4$. It is interesting to see that for even an infinitesimal $\phi>\pi/4$, the SOC needed to restore ${\mathcal{PT}}$ symmetry is $\lambda\sim {\mathcal{O}}(1)$. This is in contrast to the three site model where the line of exceptional points was given by $\phi-\pi/4\propto \lambda^2$ for $\lambda\rightarrow 0$. The reason for the discrepancy is that the full model has a conduction band, and we find that the minimum spin-orbit coupling needed to restore ${\mathcal{PT}}$ symmetry is of the order of bandwidth, which is ${\mathcal{O}}(1)$ in the present case, and was zero in the three-site model.

\subsection{Spectral sum rule in the \texorpdfstring{$U=0$}{Lg}, non-interacting case}

Since the model is PT-symmetric, and the eigenvalues of the Hamiltonian are real in a finite range of the parameter space, the time evolution of operators will be unitary, for $\phi < \phi_{\scriptscriptstyle{EP}}$, where the latter represents the exceptional point. Hence, we can use the equation of motion (EoM) method to find the retarded Green's functions ($G^A_B(t,t^\prime)=-i
\langle\Psi_G^L| \{A_H(t),B^\dagger_H(t^\prime)|\Psi_G^R\rangle\theta(t-t^\prime)$) in the unbroken ${\mathcal{PT}}-$symmetry regime\cite{moiseyev2011non}. 
 The Green's function is defined with respect to the left and right (L/R) eigenstates of the full Hamiltonian. 
The equations of motion are given by
\begin{equation}
    \begin{split}
\omega^+G^A_B(\omega) &= \langle \{A,B\}\rangle +
G^{[A,H]_-}_B(\omega) \\
&= \langle \{A,B\}\rangle +
G_{[H,B]_-}^A(\omega)
    \end{split}
\end{equation}
In order to gain insight into the interplay of PT-symmetry and SOC, we investigate the non-interacting case ($U=0$).
We first find the retarded Green's function
for the dot operators in the chiral basis,
i.e $G^{dd}_{hh}$. The following equation is obtained for the dot Green's functions,
\begin{equation}
    G^{dd}_{hh}= \left[\omega^+ -\epsilon_d -
    \Gamma_h(\omega)\right]^{-1}
    \label{eq:nintgd}
\end{equation}
where the hybridization function is given as 
\begin{equation}
    \Gamma_h(\omega)=\sum_{k}\frac{X_{kLh}^2 + X_{kRh}^2}{\omega^+ - \tilde{\epsilon}_{kh}} 
        \label{eq:hyb}
\end{equation}
 Using $X_{Lk+}=X_{Lk-}= |X_k|e^{i\phi_k}$, so that $X_{kR+}=-|X_k|e^{-i\phi_k}=-X_{kR-}$, the diagonal dot Green's functions are obtained as (for $U=0$):
\begin{equation}
\label{eq:gdot}
   G^{dd}_{hh}(\omega)=
    \frac{1}{\omega^+ - \epsilon_d - 2\sum_{k}
    \frac{|X_k|^2\cos(2\phi_k)}{\omega^+ - \tilde{\epsilon}_{kh}}}
\end{equation}
The Green's function is causal, as long as either (i) $\phi_k\le \pi/4,\;\forall\, k$ or (ii) if the imaginary part of hybridization is zero. The spectral density is guaranteed to be positive definite in this regime. Interestingly, although the Hamiltonian is non-Hermitian for $\phi_k\ne 0$, the Green's function, being exactly the same as for a single impurity Anderson model with a renormalized hybridization ($V_k^2 \rightarrow 2|X_k|^2\cos(2\phi_k)$), is fully causal, and has the same analytic structure of the usual Green's functions. 
We also see that the eigenvalue equation~(\ref{eq:nieig}) is identical to the equation obtained for the zeroes of the denominator of the Green's function (\ref{eq:gdot}). This is not surprising, since the non-interacting Green's function is given by 
$\hat{G} = \left[\omega^+\mathbf{1} - {\hat{H}}\right]^{-1}$.

Taking $X_k=X_0$ and $\phi_k=\phi$ to be independent of $k$, the $k$-sum in the hybridization can be converted to a density of states integral, which is then given by
\begin{equation}
    \Gamma_h(\omega)=\Gamma_0\cos(2\phi_0)H_T[\omega^+ -h\lambda]\,,
    \label{eq:defhilb}
\end{equation}
 where $\Gamma_0=2X_0^2$ and $H_T[z]$ is the Hilbert transform defined by 
\begin{equation}
  H_T[z]=  \int^\infty_{-\infty} \,d\epsilon\, \frac{\rho_0(\epsilon)}{z - \epsilon} \,.
\end{equation}
The exact diagonalization calculation required us to choose energy values, $\epsilon_k$, distributed in a certain way, and for convenience, we chose a uniform distribution. The density of states accordingly for the Green's function calculation has been chosen to be a flat band, namely
$\rho_0(\epsilon)=
\sum_k\delta(\epsilon - \tilde{\epsilon}_{kh}) = \theta(D-|\epsilon_{kh}|)/(2D)$, for which the Hilbert transform may be obtained in a straightforward way as:
\begin{equation}
    H_{Th}[\omega^+] = \frac{1}{2D}\ln\left|\frac{\omega-h\lambda+D}{\omega-h\lambda-D}\right| -i\frac{\pi}{2D}\theta\left(D-|\omega-h\lambda|\right)
\end{equation}

\begin{figure}[t]
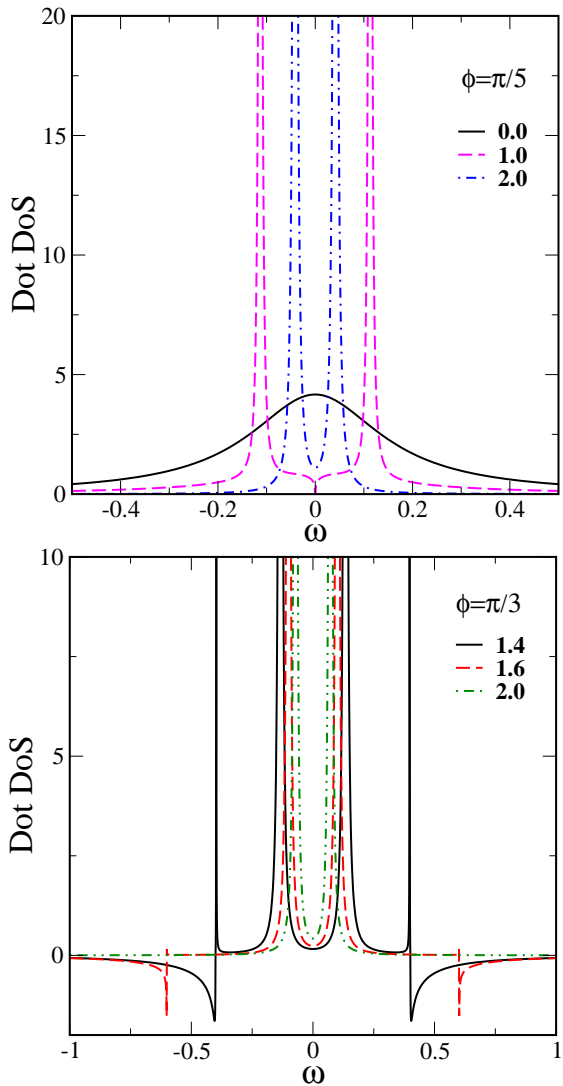

    \centering
    \includegraphics[scale=0.55,trim=0 0 0 0, clip]{dos_varying_lambda_phi_lt_Phic.eps}
    \includegraphics[scale=0.55,trim=0 0 0 0, clip]{dos_varying_lambda_phi_gt_Phic.eps}
     \caption{Dot density of states (DoS) as a function of frequency for various values of SOC (legends are the values of $\lambda$) and  $\phi=\pi/5$ (top panel) and $\phi=\pi/3$ (bottom panel). In the top panel, the system does not violate ${\mathcal{PT}}$-symmetry for $\phi=\pi/5$ for any $\lambda$. The bottom panel is for $\phi=\pi/3$, for which $\lambda=1.5$ represents the SOC value beyond which ${\mathcal{PT}}$-symmetry is restored. }
      \label{fig:dos_l_phi0}
\end{figure}

The spectral function is given by
\begin{equation}
    D_h(\omega)=-\frac{1}{\pi}{\rm Im} G^{dd}_{hh} = -\frac{1}{\pi}{\rm Im}\left(\omega^+ - \epsilon_d - \Gamma_h(\omega)\right)^{-1}\,.
\end{equation}
For a representative set of parameters $\Gamma_0=1/4$, we show the spectral function of the chiral index summed Green's function ($G_d=0.5\sum_h G^{dd}_{hh}$) in figure~\ref{fig:dos_l_phi0}.
The numbers mentioned in the legends are values of the SOC($\lambda$).
Both panels shows the dot density of states (DoS) in the non-Hermitian case of  $\phi=0.2\pi < \pi/4 $ (top) and $\phi=\pi/3 > \pi/4$ (bottom). The top panel shows that for $\lambda=0$, the DoS is a lorentzian, as expected, while for higher $\lambda$, the DoS splits into two peaks. These peaks grow in intensity, while becoming narrower as $\lambda \gg 1$. In fact, in the latter limit, it is easy to show that the DoS reduces to just two poles at $\omega=\pm 2\Gamma_0\cos 2\phi/|\lambda|$.  We see that for any value of $\lambda$ in the top panel, the DoS preserves the spectral sum rule, and is hence causal.  The bottom panel shows the DoS for $\phi=\pi/3$, which is greater than $\pi/4$, so for $\lambda=0$, the DoS should be expected to be acausal. Indeed, we see that for $\lambda=1.4$ (and all lower values), the DOS is negative and acausal, but for all $\lambda > 1.5$, we recover causality, in the sense that the integrated spectral weight or the spectral norm is one. But as the figure shows, the DoS does become negative over a finite frequency range, albeit, the negative weight is compensated by the positive part, preserving the total spectral norm. The density of states being negative is of course not physical in a conventional Hermitian picture; but for a non-Hermitian system considered here, such a result can be speculated to imply states that are not stationary and are either lossy or amplifying. We note that further studies are required for finding the correct interpretation of negative density of states. The inference from the above investigation is that higher $\lambda$ values restore ${\mathcal{PT}}$-symmetry
that was broken spontaneously at lower $\lambda$, and hence SOC protects 
${\mathcal{PT}}$-symmetry.

In the hermitian case, the spectral function is positive-definite and causal, i.e, normalized to one ($\int\,d\omega D_h(\omega) = 1$). We will explore 
the violation of causality by computing the deviation of the norm from unity, as a function of $\phi$ and $\lambda$.
 \begin{figure}[t]
    \centering
\includegraphics[scale=0.55,trim=0 0 0 0, clip]{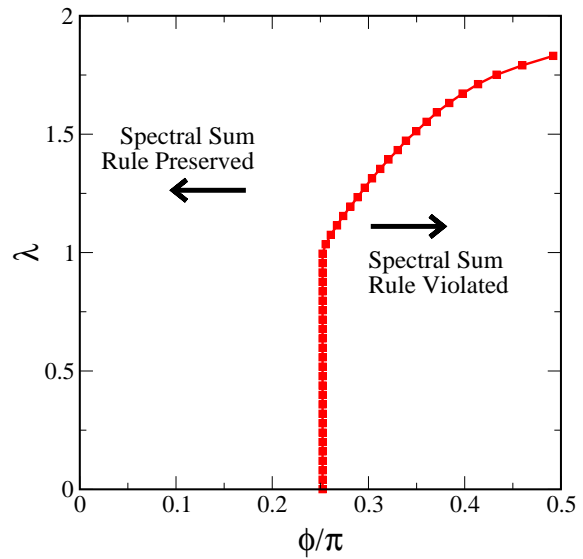}
     \caption{The $U=0$ Phase diagram in the $\phi-\lambda$ plane found through the norm violation condition of the Green's function. The solid symbols represent the line of exceptional points, where the spectral sum rule gets violated and the norm deviates from unity.}
      \label{fig:phase_nint_gf}
\end{figure}
The most interesting feature about this phase diagram is that the hybridization, being proportional to $\cos(2\phi)$ has an acausal imaginary part beyond $\phi>\pi/4$, nevertheless, the spectral function sum rule is not violated above a certain value of the spin-orbit coupling, $\lambda$. 
The critical $\lambda$ for $\phi=\pi/4$ is found to be equal to the effective bandwidth of the conduction band. Since $\lambda$ shifts the centre of the conduction band away from $\omega=0$ for each 'chiral' index, this implies that if  the imaginary part of hybridization is either vanishingly small or negative definite (causal) at $\omega=0$, the spectral function norm is preserved. Thus, as found through exact diagonalization, the spin-orbit coupling stabilizes ${\mathcal{PT}}-$symmetry. 

We note that such a conclusion is sensitive to the choice of the conduction band density of states. For example, the choice of an infinitely wide flat band will result in the exceptional point being $\phi_{EP}=\pi/4$ or $g_{EP}=1$ for any value of the spin-orbit coupling, $\lambda$. We will find the trace of the Hamiltonian in the next section, and identify the violation of ${\mathcal{PT}}$-symmetry
from the dependence of total energy on $\phi$ and $\lambda$.

\subsection{Total energy of the causal states (\texorpdfstring{$U=0$)}{Lg}}
A violation of ${\mathcal{PT}}$-symmetry introduces complex conjugate eigenvalues into the eigenspectrum. Thus, if we measure the energy of the system as the trace over the occupied states of the real part of the eigen-spectrum, we should observe a discontinuity when one or more pairs of real eigenvalues becomes complex. Such a discontinuity provides an alternate measure of the exceptional point. 

\begin{figure}[t]
    \centering
    \includegraphics[scale=0.55,trim=0 0 0 0, clip]{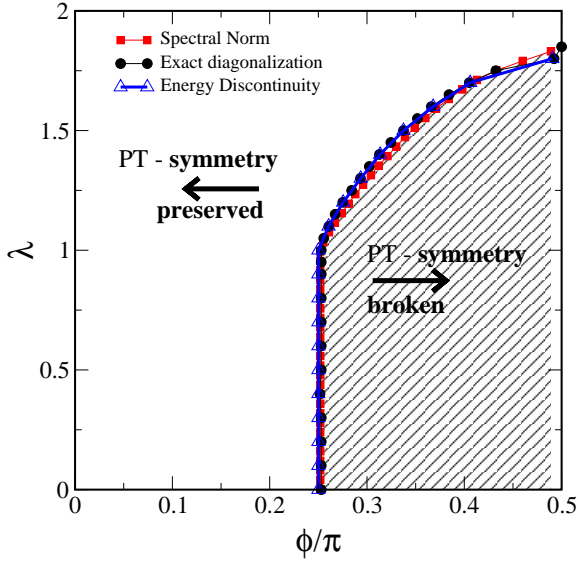}
 \caption{The phase diagram in the $\phi-\lambda$ plane, computed through an identfication of the energy discontinuity (see text for details)  showing the line of exceptional points (triangles). The shaded region represents the spontaneously broken ${\mathcal{PT}}$-symmetry broken regime, and the unshaded region represents the ${\mathcal{PT}}$-symmetry protected regime. The circles and squares represent the exceptional points derived through exact diagonalization and the spectral sum rule violation respectively. All the three conditions for determining exceptional points agree very well.}
      \label{fig:ep_toten}
\end{figure}
Figure~\ref{fig:ep_toten} shows the phase diagram computed through the energy discontinuity superimposed on the phase diagrams obtained through exact diagonalization and the spectrum sum rule violation.
The three criteria for finding the exceptional points match very well. Thus, we establish that {\em the ${\mathcal{PT}}$-violation as seen by the exact diagonalization of the Hamiltonian is reflected precisely by causality violation of the Green's function, and by the energy discontinuity condition.}

Until now, we have restricted ourselves to the non-interacting case ($U=0$). In the following subsection, we will investigate the other extreme, i.e the $U\rightarrow\infty$ limit using the slave boson formalism.

\subsection{Slave boson mean-field solution}
The Hamiltonian in equation~\ref{eq:Canderson} may be rewritten in the $U\rightarrow\infty$ case using Coleman bosons \cite{newns1987mean} as 
\begin{equation}
   \begin{split}
    H &= H_0  \\
 & +\sum_{k\eta h} X_{k\eta h}\left( c^{\dagger}_{k\eta h}b_{\eta}^{\phantom{\dagger}} d_{h}^{\phantom{\dagger}}+ {\rm h.c.}\right)\\
 &+\epsilon_d \sum_h n_{dh} +\zeta\left(\sum_\eta b^\dagger_{\eta} b^{\phantom{\dagger}}_\eta + \epsilon_d \sum_{h}n_{dh} -1\right)
\end{split}
    \label{eq:hamsbeqns}
\end{equation}
where $H_0=\sum_{kh\eta}\tilde{\epsilon}_{kh}^{\phantom{\dagger}} c_{kh\eta}^\dagger c_{kh\eta}^{\phantom{\dagger}}$, and $\zeta$ is the Lagrange multipler which enforces the 
constraint that the total filling (fermions + bosons) is one.
Note that the two channels, namely $L$ and $R$ have been associated with two different bosons as a general possibility. 
Hence, with the mean-field approximation,
$\langle b^{\dagger}_{L} \rangle=\langle b_{L} \rangle=re^{i\theta}$ and $\langle b^{\dagger}_{R} \rangle=\langle b_{R} \rangle=re^{-i\theta}$ implying that the total mean boson number will be $\langle b^{\dagger}_{L}b_{L}+b^{\dagger}_{R}b_{R}
\rangle=2r^2\cos(2\theta)$, the mean field Hamiltonian becomes
\begin{equation}
   \begin{split}
    H_{MF} &= H_0 
  +\sum_{k\eta h} \tilde{X}_{k\eta h}\left[ c^{\dagger}_{k\eta h} d_{h}^{\phantom{\dagger}}
  +  {\rm h.c.}\right]\\
 &+\tilde{\epsilon}_d \sum_h n_{dh} + \zeta\left(2r^2\cos(2\theta)  -1\right)
\end{split}
    \label{eq:hamsb}
\end{equation}
where  $\tilde{X}_{kL h}=re^{i\theta} X_{kLh}=
\sqrt{2}r|X_k|e^{i(\phi+\theta)}$, $\tilde{X}_{kR h}={\bar{h}}re^{-i\theta} X_{kRh}=
\sqrt{2}{\bar{h}}r|X_k|e^{-i(\phi+\theta)}$ and  $\tilde{\epsilon}_d=\epsilon_d + \zeta$.
The parameters $r,\zeta$ and $\theta$ may be found self-consistently by minimizing $\langle H_{MF}\rangle$.  Since the slave boson mean-field Hamiltonian has exactly the same form as the non-interacting Hamiltonian, equation~\ref{eq:Canderson}, we will find the expression for the total energy in the $U=0$ case, and generalize it to the slave-boson case.

For finding the expectation value of the Hamiltonian in equation~\ref{eq:Canderson}, we need Green's functions other than the one computed before (equation~\ref{eq:gdot}). These are listed below:
\begin{align}
    G^{k\eta h}_{k\eta h} &= \frac{1}{\omega^+ -
    \tilde{\epsilon}_{kh}} + \frac{X_{k\eta h}^2}{(\omega^+ -
    \tilde{\epsilon}_{kh})^2} G^{dh}_{dh}\\
    G_{k\eta h}^{dh} &= G^{k\eta h}_{dh} = \frac{X_{k\eta h}}{\omega^+ - \tilde{\epsilon}_{kh}}G^{dh}_{dh}
\end{align}
Using these to find the expectation value of the Hamiltonian,
equation~\ref{eq:Canderson}, we get
\begin{equation}
    E_{tot} = \int^\infty_{-\infty}\,d\omega\,f(\omega) \tilde{D}(\omega)
\end{equation}
where 
\begin{equation}
\begin{split}
    &\tilde{D}(\omega)=-\frac{1}{\pi}{\rm Im} \sum_h \tilde{G}_h(\omega)\\
    &=-\frac{1}{\pi}{\rm Im}\Bigg[\sum_{kh\eta} \left(\tilde{\epsilon}_{kh} G^{kh\eta}_{kh\eta}
    +2X_{kh\eta}G^{dh}_{k\eta h}\right)
    + \epsilon_d \sum_{h} G^{dh}_{dh}\Bigg]
    \end{split}
\end{equation}
and,
\begin{equation}
    \begin{split}
        &\tilde{G}_h(\omega) = \left(-\sum_{k\eta}\right) + \left(\omega^+\sum_{k\eta}\frac{1}{\omega^+ -\tilde{\epsilon}_{kh} }\right) \\
    + & G^{dh}_{dh}\left(\epsilon_d + \sum_{k\eta}
    \frac{X^2_{kh\eta}}{\omega^+ -\tilde{\epsilon}_{kh}} + \omega^+\sum_{k\eta} \frac{X^2_{kh\eta}}{(\omega^+ -\tilde{\epsilon}_{kh})^2}\right)
    \end{split}
\end{equation}
which when simplified yields
\begin{equation}
\begin{split}
    \tilde{D}_h(\omega)&
  =\omega D_{c0}(\omega-h\lambda) + \epsilon_d D_{d0h}(\omega) \\
  &-\frac{1}{\pi}{\rm Im}\Bigg[
   G^{dh}_{dh}\left(
  \Gamma_h- \omega\frac{d\Gamma_h}{d\omega}\right)\Bigg]
\end{split}
\end{equation}
The first term contributes to the conduction electron energy, and depends on $\lambda$ as follows:
\begin{equation}
\begin{split}
    &E_{c0}=\sum_h\int^0_{-\infty} \omega D_{c0}(\omega -h\lambda)\,d\omega \\
    & =2 \int^0_{-\infty} \omega D_{c0}(\omega)\,d\omega +
    2\int^\lambda_{0}( \omega-\lambda) D_{c0}(\omega)\,d\omega \\
    &= E_0 + 2\int^\lambda_{0}( \omega-\lambda) D_{c0}(\omega)\,d\omega\,
\end{split}
\end{equation}
where $E_0$ is independent of $\lambda$, and we have assumed that $D_{c0}(\omega)$ is symmetric about $\omega=0$. The second term yields a contribution proportional to the dot occupancy. Thus
\begin{equation}
\begin{split}
    E_{tot} &-E_0=  2\int^\lambda_{0}( \omega-\lambda) D_{c0}(\omega)\,d\omega + \epsilon_d \sum_h n_{dh0}\\
    &-\sum_h\int^0_{-\infty}\,d\omega\,\frac{1}{\pi}{\rm Im}\Bigg[
   G^{dh}_{dh}\left(
  \Gamma_h- \omega\frac{d\Gamma_h}{d\omega}\right)\Bigg]
\end{split}
\label{eq:toten}
\end{equation}

A Gaussian conduction band is chosen for convenience as $\rho_0(\epsilon)=
\sum_k\delta(\epsilon - \tilde{\epsilon}_k) = \exp(-\epsilon^2/2t_*^2)/\sqrt(\pi)t_*$ (with $t_*=1$ as the unit of energy), for which the Hilbert transform may be written in terms of the Faddeeva function, $w(z)$ as 
$H_T[z] = (-i\sqrt{\pi}/t_*)w(z/\sqrt{2}t_*)$ if ${\rm Im}(z) > 0$.
With a Gaussian DoS, the hybridization function, $\Gamma_h(\omega)$ is given
by (equation~\ref{eq:defhilb})
\begin{equation}
\begin{split}
    \Gamma_h(\omega)& = \Gamma_0\cos(2\phi_0)H_T[z_h]\\
    &=\Gamma_0\cos(2\phi)\left(-is\sqrt{\pi}\exp(-z_h^2){\rm erfc}(-isz_h)\right)
\end{split}
\end{equation}
where $z_h=\omega^+-h\lambda$, $s={\rm sgn}({\rm Im}z_h)=+1$,
and ${\rm erfc}(z)$ is the complementary error function.
The derivative of the hybridization function is given by
\begin{equation}
  \begin{split}
\frac{d\Gamma_h}{d\omega} &=  \Gamma_0\cos(2\phi) \frac{dH_T(z)}{dz} \\
  &= 2\Gamma_0\cos(2\phi)\left(1-zH_T[z]\right)
  \end{split}
\end{equation}
The total energy expression~\ref{eq:toten} shows that the first term is the conduction electron contribution, and the second and third terms are the contributions due to the dot and the hybridization respectively. If $\lambda=0$, and $\phi=\pi/4$, then for $\epsilon_d=0$, the total energy is just zero. In fact, the third term, being proportional to $\cos(2\phi)$ will yield zero for any $\lambda$ and $\epsilon_d$ at $\phi=\pi/4$.

The slave-boson mean-field Hamiltonian may be treated exactly as the non-interacting limit, and 
the average energy $\tilde{E}_{\rm tot}=\langle H_{MF}\rangle$ is obtained as
\begin{equation}
\begin{split}
    \tilde{E}_{tot} &-E_0=  2\int^\lambda_{0}( \omega-\lambda) D_{c0}(\omega)\,d\omega + \tilde{\epsilon}_d \sum_h \tilde{n}_{dh0}\\
    &-\sum_h\int^0_{-\infty}\,d\omega\,\frac{1}{\pi}{\rm Im}\Bigg[
   \tilde{G}^{dh}_{dh}\left(
  \tilde{\Gamma}_h- \omega\frac{d\tilde{\Gamma}_h}{d\omega}\right)\Bigg]\\
 &+ \zeta\left(2r^2\cos(2\theta)  -1\right)
\end{split}
\label{eq:esbmf}
\end{equation}
where the renormalized dot Green's functions and hybridizations are
given by
\begin{align}
\tilde{G}^{dh}_{dh}(\omega)&=\left[\omega^+ -\tilde{\epsilon}_d -\tilde{\Gamma}_h(\omega)\right]^{-1}\,,\\
\tilde{\Gamma}_h(\omega) & = 4r^2|X_0|^2\cos(2\phi+2\theta)\Delta_h(\omega)\,, 
\end{align}
and the bare hybridization functions $\Delta_h(\omega)$ are given in equation~\ref{eq:defhilb}. We observe that the first term in equation~\ref{eq:esbmf} does not depend on $r,\zeta$ or $\theta$, and hence will not affect the minimization. 

\begin{figure}[h!]
    \centering{
    \includegraphics[scale=0.55,trim=0 0 0 0, clip]{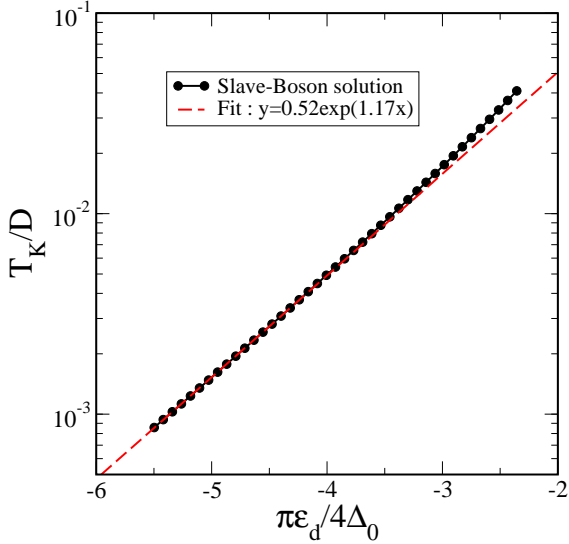}}
    \caption{$T_K$ (black, solid)
and the exponential fit(red-dashed)  as a function of scaled dot-orbital energy, $\pi\epsilon_d/4\Delta_0$ for $\lambda=0$ and $\phi=0$. }
      \label{fig:tk_sb_gauss}
\end{figure}

\begin{figure}[h!]
    \centering
    \includegraphics[scale=0.55,trim=0 0 0 0, clip]{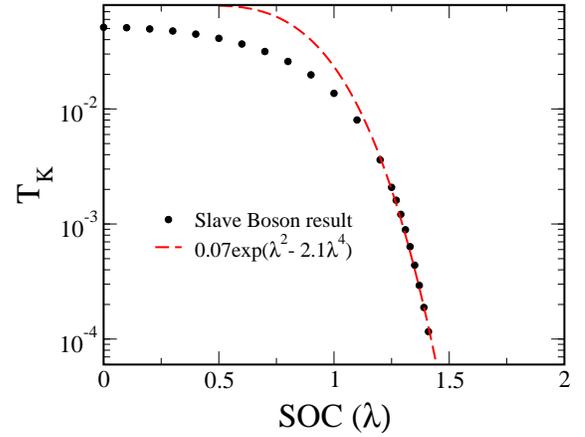}
     \caption{$T_K$ (black, solid)
and the exponential fit(red-dashed)  as a function of spin-orbit coupling ($\lambda$). }
      \label{fig:tk_sb_soc}
\end{figure}

In order to investigate the effect of finite bandwidth and SOC in the strong coupling limit, we go back to equation~\ref{eq:esbmf}, and minimize the total energy for a Gaussian band and for finite SOC. We will find the equations for the determination of $r^2, \tilde{\epsilon}_d$ and $\theta$, and solve them numerically. The derivatives of the total energy are
\begin{align}
    \frac{\partial \tilde{E}_{tot}}{\partial \zeta}& =
   {\mathcal{A}}(\tilde{\epsilon}_d, r^2,\theta) + 2r^2\cos2\theta -1  \\
  \frac{\partial \tilde{E}_{tot}}{\partial r^2}& = \frac{1}{r^2} {\mathcal{B}}(\tilde{\epsilon}_d, r^2,\theta) +2\zeta\cos2\theta\\
  \frac{\partial \tilde{E}_{tot}}{\partial \theta}& = -2\left[\tan(2(\phi+\theta)){\mathcal{B}}(\tilde{\epsilon}_d, r^2,\theta) +2\zeta r^2\sin2\theta\right]\,, 
\end{align}
where
\begin{equation}
\begin{split}
{\mathcal{A}}(\tilde{\epsilon}_d, r^2,\theta) =&-\frac{1}{\pi}{\rm Im} \int^0_{-\infty} \sum_h {\mathcal{F}}_h(\omega)\,d\omega \\
{\mathcal{B}}(\tilde{\epsilon}_d, r^2,\theta) =&-\frac{1}{\pi}{\rm Im} \int^0_{-\infty} \sum_h {\mathcal{G}}_h(\omega)\,d\omega
\end{split}
\end{equation}
and
\begin{equation}
\begin{split}
{\mathcal{F}}_h(\omega) =& \left(\tilde{G}^{dh}_{dh}\right)^2 \left(\tilde{\epsilon}_d + \tilde{\Gamma}_h -\omega \frac{d \tilde{\Gamma}_h}{d\omega}\right) + \tilde{G}^{dh}_{dh}\\
{\mathcal{G}}_h(\omega) =& \left(\tilde{G}^{dh}_{dh}\right)^2
\tilde{\Gamma}_h \left(\tilde{\epsilon}_d + \tilde{\Gamma}_h -\omega \frac{d \tilde{\Gamma}_h}{d\omega}\right) \\
&+ \tilde{G}^{dh}_{dh} \left( \tilde{\Gamma}_h -\omega \frac{d \tilde{\Gamma}_h}{d\omega} \right)
\end{split}
\end{equation}
Using the above expressions, and the definitions
we get the following self-consistent nonlinear equations:
\begin{align}
  {\mathcal{A}}(\tilde{\epsilon}_d, r^2,\theta) & = 1-2r^2\cos(2\theta)\label{eq:sbged} \\
  {\mathcal{B}}(\tilde{\epsilon}_d, r^2,\theta) & = -2\zeta r^2\cos(2\theta) \label{eq:sbgr}\\
 \tan(2\phi+2\theta){\mathcal{B}}(\tilde{\epsilon}_d, r^2,\theta) & = -2\zeta r^2\sin(2\theta) \label{eq:sbgtheta}\,.
\end{align}
 Since equations~\ref{eq:sbgr} and \ref{eq:sbgtheta} yield the result that $\theta\neq 0$ only if $\phi=0$, we conclude that the slave-boson equations do not renormalize the non-Hermitian coupling strength, which is non-zero only if $\phi\ne 0$. Thus, we restrict ourselves to $\theta=0$, and solve equations ~\ref{eq:sbged} and \ref{eq:sbgr} to determine $\tilde{\epsilon}_d = \epsilon_d+\zeta$ and $r^2$. Again, we choose a Gaussian density of states for the bare conduction band. 
The numerical solution proceeds with the choice of a parameter, $a=r^2/\tilde{\epsilon}_d$, and using this, the equation~\ref{eq:sbged} can be transformed to a single variable non-linear equation, and hence can be solved easily.
Given both $r^2$ and $\tilde{\epsilon}_d$, we can substitute in equation~\ref{eq:sbgr} and get $\zeta$, and hence $\epsilon_d$. For vanishing SOC ($\lambda\rightarrow 0$), we expect to find a Kondo scale that has a dependence similar to the conventional Hermitian case. Figure~\ref{fig:tk_sb_gauss} shows that the Kondo scale, $T_K/D=r^2\sqrt{{\Delta}_0^2 + \tilde{\epsilon}_d^2}/D$, found with the Gaussian density of states, is indeed exponentially dependent on $\pi\epsilon_d/\Delta_0$, but with an exponent, that is slightly different than the one obtained for the flat band case. 
With increasing SOC, since the hybridization value at $\omega=0$ decreases, we may expect that the Kondo scale should also decrease. A countereffect is provided by a concomitant increase in the total bandwidth. However, since $\lambda$ affects $\Delta_h(\omega=0)$ exponentially (Gaussian DoS), the Kondo scale decreases exponentially with an increase in SOC as shown in figure~\ref{fig:tk_sb_soc}. This finding must be contrasted with that of the flat band case\cite{quasilin,small}, where, since the $\Delta_h(\omega=0)$ does not change with varying $\lambda$, the bandwidth becomes the only controlling parameter, and the scale increases linearly with increasing SOC in the flat band case. 

Now, we investigate the variation of the scale with increasing non-hermitian strength for fixed SOC and $\epsilon_d$ in the strong coupling regime. To recapitulate the results from the non-interacting case, we had found that for $\lambda=0$, the exceptional point was at $\phi_{EP}=\pi/4$. And increasing $\lambda$, increased the $\phi_{EP}$ to beyond $\pi/4$, showing that SOC stabilized ${\mathcal{PT}}$-symmetry for $U=0$. In the
zero bandwidth, three-site model, we found that the exceptional point ($g_{\scriptscriptstyle{EP}}>1$) and the quantum critical point ($g_c=1$) become distinct in the $U-\lambda$ plane except on the $\lambda=U/2$ line,
where $g_{\scriptscriptstyle{EP}}=g_c=1$. With the slave-boson calculation, we will be able to extract the Kondo scale, $T_K$, and hence the quantum critical point will be identified through the vanishing of $T_K$. However, 
we may not be able to identify the exceptional point since the zero bandwidth case indicates that ${\mathcal{PT}}$-symmetry breaks spontaneously at a non-hermitian coupling, $g\geq 1$, which is always greater than or equal to the QCP ($g=1$); while the slave boson mean-field vanishes at the QCP, and the theory may not even be valid beyond the QCP. Now, we discuss the slave-boson results. 

\begin{figure}[h!]
    \centering
    \includegraphics[scale=0.55,trim=0 0 0 0, clip]{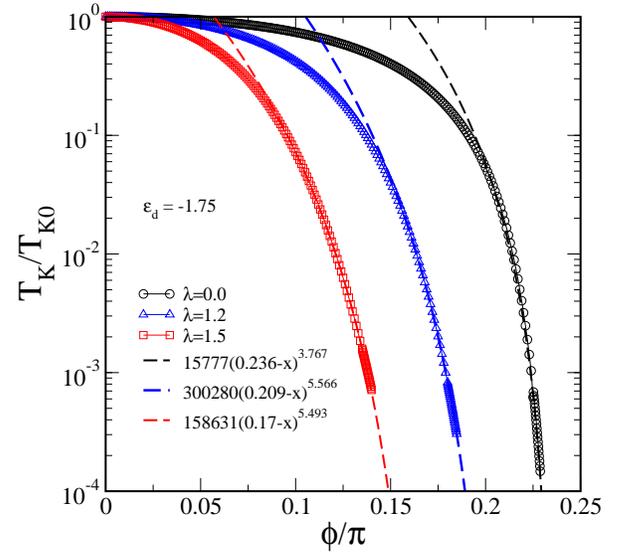}
     \caption{$T_K/T_0$ for $\lambda=0$ (black circles) and $\lambda=1.5$ (red squares)
and the respective fits (green-dashed and 
blue-dashed) for $\epsilon_d=-1.75$. }
      \label{fig:tk_phi0_various_soc}
\end{figure}

In figure~\ref{fig:tk_phi0_various_soc}, we show $T_K/T_{K0}$ {\it vs.} $\phi$ for $\epsilon_d=-1.75$ and $\lambda=0, 1.2$ and $1.5$. The scale decreases sharply with increasing $\phi$, and is seen to vanish at a critical $\phi$ (as seen by the power law fits). 
The critical values of $\phi$ and the non-Hermitian coupling $g=\tan(\phi)$ for various $\lambda$ are given in table~\ref{tb:phic}. It is observed that the critical coupling decreases sharply with increasing $\lambda$, and at some value of $\lambda=\lambda_c$, the critical $\phi$ will vanish, which implies that the model will not have a Kondo screened state for any finite value of the non-Hermitian coupling if $\lambda > \lambda_c$.
The results shown in table~\ref{tb:phic} also consolidate the inference that interactions and SOC cooperate in reducing the value of the quantum critical non-Hermitian strength.
\begin{center}
\begin{table}
\begin{tabular}{ || m{5em} | m{5em}| m{8em} || }
 \hline
 $\lambda$ & $\phi_c$ & $g_c=\tan(\phi_c)$ \\ [0.5ex] 
 \hline\hline
 0.0 & $0.236\pi$ & 0.916 \\ 
 \hline
 1.2 & $0.209\pi$ & 0.771  \\
 \hline
 1.5 & $0.17\pi$ & 0.591  \\
 \hline
\end{tabular}
\caption{Table showing the results of the fits of $T_K$ {\it vs.} $\phi$ of figure~\ref{fig:tk_phi0_various_soc}, from which the critical $\phi$ has been obtained.}
\label{tb:phic}
\end{table}
\end{center}

\begin{figure}[h!]
    \centering
    \includegraphics[scale=0.55,trim=0 0 0 0, clip]{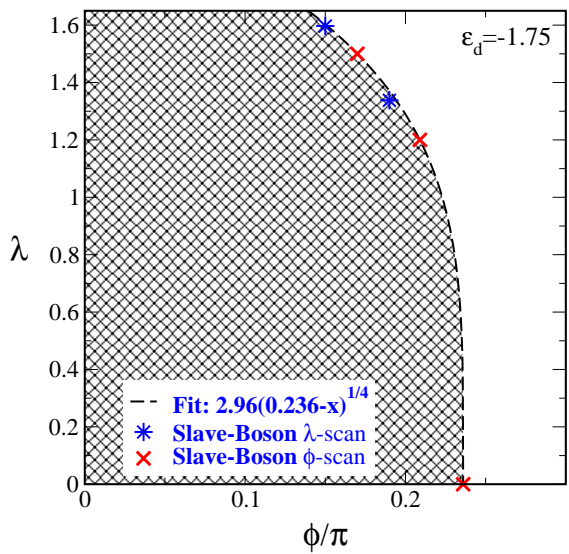}
     \caption{The shaded region represents a parameter regime, where a finite Kondo scale is found. The slave-boson results for the vanishing of $T_K$ are shown by red crosses ($\phi$-scan), and blue stars ($\lambda$-scan) and the dashed line represents a fit, and is a guide to the eye.  }
      \label{fig:SOC_Phi_Phase_Diag}
\end{figure}

Using the above results in figure ~\ref{fig:tk_phi0_various_soc}, we can draw a phase diagram in the $\phi-\lambda$ plane. The shaded region in figure~\ref{fig:SOC_Phi_Phase_Diag} represents the parameter regime where the Kondo scale is finite, while the dashed line is where the scale vanishes. The red crosses are the actual slave-boson results found through the analysis shown in figure~\ref{fig:tk_phi0_various_soc}, where for a fixed $\lambda$, we have found the $T_K$ {\it vs} $\phi$. This will be called $\phi$-scan. In order to validate the phase diagram, we have carried out $\lambda$-scans for fixed $\phi$, and the blue stars shown in figure~\ref{fig:SOC_Phi_Phase_Diag} are found to lie on the same curve as the red crosses. The fit shows that the critical $\lambda_c$ discussed above is given by
$\lambda_c=2.96(0.236)^{1/4} = 2.06$, beyond which the model does not support the Kondo screened state for any finite $\phi$.

\section{Discussion and Conclusions}

In this work, we have considered the interplay of interactions, Rashba spin-orbit coupling and non-Hermitian coupling to the baths on the Kondo effect and preservation or violation of ${\mathcal{PT}}$-symmetry. We begin with a derivation of the model of an interacting quantum dot hybridizing through non-Hermitian couplings with non-interacting leads having Rashba spin-orbit coupling using Lindbladian dynamics.
A simplification of the full model in terms of a zero bandwidth, three-site model is considered, which yields a wealth of information including the demonstration of ${\mathcal{PT}}$-symmetry, the dependence of quantum critical points and exceptional points on Rashba spin-orbit coupling and interactions etc.
Our analysis shows that the exceptional point at $g_{\scriptscriptstyle{EP}}=1$ in the non-interacting case coincides with the Kondo destruction critical point, $g_c$ for all $\lambda=U/2, U\neq 0$, but the two bifurcate significantly everywhere else in the $U-\lambda$ phase diagram. The quantum critical point remains at $g_c=1$ for all $U\neq 0$ and $\lambda$. The phase diagram of this simple, three-site system shows that $U$ and $\lambda$ protect ${\mathcal{PT}}$ symmetry in the $U/\lambda \gg 1$ and $U/\lambda \ll 1$ regime for any strength of the non-Hermitian coupling, but in the neighbourhood of $\lambda=U/2$, the exceptional point occurs at a finite  coupling strength. 

A detailed analysis of the full, finite bandwidth model, in the non-interacting case using exact diagonalization (ED), Green's functions and Hamiltonian trace calculation is used to establish that exceptional points may be deduced from the ED calculations, or equivalently from the violation of the spectral sum rule of the Green's functions or the energy discontinuity condition. 
Finally, the strong coupling regime is investigated using a slave-boson approach, which, by construction is valid for $U\rightarrow \infty$. 
The mean-field equations were derived through the Green's function approach. We have shown earlier (in section IV-A) that the spectral function is positive definite and norm preserving for all $\phi<\pi/4$,
while only beyond $\pi/4$, the spectral function became negative over certain frequency regions. Since the Kondo destruction critical point has been found for all $\lambda$ to lie at $\phi_c<\pi/4$, the derived mean-field equations are valid.
The exceptional points could not be found within this approach since the solution to the slave-boson mean-field equations yields a vanishing boson mean field implying a Kondo destruction quantum critical point before the ${\mathcal{PT}}$-symmetry is violated. The quantum critical points for the finite bandwidth case get significantly renormalized below the $\lambda=0$ value of $g_c=1$ by the SOC. A critical value of $\lambda$ is also found beyond which the model does not support the Kondo screened singlet state for any finite value of the non-hermitian coupling.  Lourenco et al~\cite{loure} had considered a real-space non-hermitian model that was similar to what we have considered, and through RG, the authors had found that the exceptional point and the critical point coincide at $g_c=1$, and the RG flow does not renormalize the critical point. This is of course, contrary to our findings, and the discrepancy could be due to the subtle differences between their model and our model. We are working on trying to understand the origin of these discrepancies through RG based approaches.
Further, the solution of the slave boson equations beyond the Kondo destruction critical point are being attempted. We believe that the slave boson mean field could take on negative or complex values implying complex values of the boson mean fields.

\acknowledgments

We acknowledge funding from JNCASR, SERB (EMR/2017/005398), and the national supercomputing mission (DST/NSM/R\&D\_HPC\_Applications/2021/26).
We acknowledge discussions with Siddhartha Lal and Anirudha Mirmira. VMK and A.Gupta acknowledge the Non-Hermitian Physics - PHHQP XVIII (ICTS) conference.

\appendix
\section{Appendix-Development of a non-Hermitian Hamiltonian}

In order to incorporate dissipation in a conventional Hermitian model, one can start with the quantum master equation\cite{ecg,linblad1,linblad2}, thereby arriving at an effective model in terms of a Lindbladian which is non-Hermitian. For spin-Hamiltonians, the Lindbladians have been constructed by Prosen and others~\cite{prosen2010spectral, prosen2012generic}. For a Hermitian Kondo model, Kawakami and co-workers~\cite{poor} utilized a similar Lindbladian approach to justify a non-Hermitian Kondo model. Lindbladians for the dissipative Bose-Hubbard model have also been considered \cite{dast2014quantum}. A unique gauge transformed hamiltonian\cite{tripathi2020mathcal} also gives rise to a $\mathcal{PT}$-symmetric, non-Hermitian 1-d Hubbard model. 

The quantum master equation is given by the following:
\begin{equation}
   \label{eq:master}
        \frac{d\tilde{\rho}}{dt}=-i[H,\tilde{\rho}]+\sum_{\sigma \sigma'}\left[\mathcal{L}_\sigma\tilde{\rho}\mathcal{L}^{\dagger}_{\sigma^{'}}-\frac{1}{2}\lbrace\mathcal{L}^{\dagger}_\sigma\mathcal{L}_{\sigma^{'}},\tilde{\rho}\rbrace
        \right]
\end{equation}
where $\mathcal{L_\sigma}$ is the ${\mathcal{PT}}$-symmetric Lindbladian obeying~\cite{prosen}, 
\begin{equation}
\label{eq:cond}
  \mathcal{P}\mathcal{T}(\mathcal{L}_\sigma) \rightarrow \bar{\sigma}\mathcal{L}_\sigma  
\end{equation}
and
$\tilde{\rho}=\mathcal{PT}|\psi\rangle\langle\psi|$ is the  ${\mathcal{PT}}$-symmetric non-Hermitian
density matrix. The above master equation can also be written in terms of diagonal generators of dynamical quantum groups by a unitary transformation which is discussed in the quantum master equation approach  to many body systems\cite{breuer2002theory}. 
The last term in eq~\ref{eq:master}  is the recycling term, which can be absorbed in the unitary part
of the single particle effective evolution~\cite{daley2014quantum}, using an effective Hamiltonian as
\begin{equation}
\label{efflin0}
        \frac{d\tilde{\rho}}{dt}=-i[H_{eff},\tilde{\rho}]+\sum_{\sigma \sigma'}\mathcal{L}_{\sigma}^{\dagger}\tilde{\rho}\mathcal{L}_{\sigma'}\,,
\end{equation}
 where the effective Hamiltonian is given by
 \begin{equation}
  H_{eff}=H+\frac{i}{2}\sum_{\sigma\sigma'}\mathcal{L}^{\dagger}_{\sigma}\mathcal{L}_{\sigma'} \,,
 \label{eq:heff}
 \end{equation}
and eq~\ref{efflin0} obeys ${\mathcal{PT}}$-symmetric Liouvillian dynamics as $\frac{d\tilde{\rho}}{dt}=\mathcal{\hat{L}}\tilde{\rho}$ as shown in Ref.\cite{prosen2012generic}, in terms of an effective Hamiltonian given by equation~\ref{eq:heff}~\cite{ecg}.
We present a possible way of deriving a non-Hermitian Hamiltonian through this formalism in the next subsection.

\section{Lindbladian Derivation}
The construction of the Lindbladian is based on the angular quantum number $j_m=m\pm\frac{1}{2}$. For simplifying notation, we drop the k(momentum) index for now, but one can also define a local operator by summing over the momentum $c_{0m\sigma} =\sum_{k} V_{k}c_{km\sigma} $. The Lindbladians below are in the $(m,\sigma)$ basis and can also be written in terms of the $(j_m,\sigma)$ as well. For these local Lindbladians we can use the angular momentum expansion $ \mathcal{L}_{\sigma} =  \frac{1}{\sqrt{2\pi }}\sum_{m=-\infty}^{\infty}\mathcal{L}_{m\sigma}\exp(-i m \theta)$ each of these can be written as $\mathcal{L}_{m\sigma}=\mathcal{L}_{j_m \sigma}$ we use the $(m,\sigma)$ bath states to represent pictorially, 
\begin{equation}
    \begin{split}
    \label{eq:sum}
        &\mathcal{L}^{\dagger}_\sigma\mathcal{L}_{\sigma'}\\
        &=\frac{1}{2\pi }\sum_{m=-\infty}^{\infty}\mathcal{L}^{\dagger}_{m\sigma}\exp(i m \theta)\sum_{m'=-\infty}^{\infty}\mathcal{L}_{m'\sigma'}\exp(-i m' \theta)\\
        &=\frac{1}{2\pi}\sum_{m,m'=0,\pm 1}e^{-i(m'-m)\theta}\mathcal{L}^\dagger_{m\sigma}\mathcal{L}_{m'\sigma'}+\text{higher m}...\\
       & =\frac{1}{2\pi} \bigg(\mathcal{L}^\dagger_{0\sigma}\mathcal{L}_{0\sigma'}+e^{-i2\theta}\mathcal{L}^\dagger_{1\sigma}\mathcal{L}_{-1\sigma'}+e^{i2\theta}\mathcal{L}^\dagger_{-1\sigma'}\mathcal{L}_{1\sigma}\\
       &+\mathcal{L}^\dag_{-1\sigma} \mathcal{L}_{-1\sigma'}+\mathcal{L}^\dag_{1\sigma}\mathcal{L}_{1\sigma'}+e^{-i\theta}\mathcal{L}^\dag_{0\sigma}\mathcal{L}_{1\sigma'}\\
       &+e^{i\theta}\mathcal{L}^\dag_{1\sigma}\mathcal{L}_{0\sigma'}+e^{i\theta}\mathcal{L}^\dag_{-1\sigma}\mathcal{L}_{0\sigma'}
       +e^{-i\theta}\mathcal{L}^\dag_{0\sigma}\mathcal{L}_{-1\sigma'}....\bigg) 
    \end{split}
\end{equation}
From the above, we can see multiplicative phase factors, but that is only for the $m\ne m'$, and it is enough to proceed with retaining only the operator structures. However, one can also start with generic expansion\ref{eq:sum} and start eliminating interaction that does not preserve the global number to get gain-loss terms.\\

We choose Lindbladian  for ground state and excited states as $\mathcal{L}_{g\sigma}=|m=0,\sigma\rangle_k+|m=\pm1,\sigma\rangle_k+|0,\sigma\rangle_d$,$\mathcal{L}_{e\sigma}=|m=0,\sigma\rangle,\text{or}|m=\pm1,\bar{\sigma}\rangle$,$\mathcal{L}_{f\downarrow}=|m=0,\downarrow\rangle$ and $\mathcal{L}_{f\uparrow}=|m=-1,\uparrow\rangle$.Here k is for bath index, and d is for impurity. This choice is on the basis low-energy physical picture of the Anderson model, where the ground state can be composed of the bath and dot states. Excited states are only bath states with lifted degeneracy due to spin-orbit interaction. Also, all Lindbladians satisfy symmetry operations $\mathcal{PT}\mathcal{L}_{m\uparrow}\to-\mathcal{L}_{m\downarrow}$, and we take purely single fermion states. To incorporate dissipation with varying number scenarios with essential inclusion of the excited states and out of them, only certain electron scattering to these states will be gain-loss balancing. There is ambiguity in choosing these states, but we can derive the various Hamiltonians connected by gauge transformations in each choice.     
\begin{equation}
    \begin{split}
    \label{eq:choice}
        &\mathcal{L}_{g\uparrow}=c_{\frac{1}{2}\uparrow}+c_{-\frac{1}{2}\uparrow}+d_{\uparrow}\\
        &\mathcal{L}_{g\downarrow}=-c_{\frac{1}{2}\downarrow}-c_{-\frac{1}{2}\downarrow}-d_{\downarrow}\\
        &\mathcal{L}_{e\uparrow}=c_{\frac{1}{2}\uparrow}\\
        &\mathcal{L}_{e\downarrow}=-c_{\frac{1}{2}\downarrow}\\
        &\mathcal{L}_{f\uparrow}=c_{-\frac{1}{2}\uparrow}\\
        &\mathcal{L}_{f\downarrow}=-c_{-\frac{1}{2}\downarrow}
    \end{split}
\end{equation}
We rewrite the sum in equation \ref{eq:sum} as $\mathcal{L}^{\dagger} _\sigma\mathcal{L}_{\sigma'}=\sum_{m,m'=g,e,f}\mathcal{L}^{\dagger} _{m\sigma}\mathcal{L}_{m'\sigma'}$ similar convention also used in the supplementary\cite{poor} for two body losses. This deviates from earlier approaches\cite{tripathi2020mathcal} like forward and backward hopping having opposite phases of each other with finite real part in 1D tight-binding models.\\

The above  Lindbladians\ref{eq:choice} give a total of 36 terms (which can be identified in \ref{eq:sum} each of the 9 terms will have implicit 4 terms with $\sigma=\pm\frac{1}{2},\sigma'=\pm\frac{1}{2}$) which are represented below pictorially. We can see several ways to derive minimal terms that preserve symmetry. We also verified, in general, that there is a possibility to derive $\mathcal{PT}$-symmetric representations if we find coefficients those commute with the total number operator (at a single particle level). Here we see it is not easy due to the angular momentum and spin indices, but this method works nicely when only the spin index is present.\\ 

\begin{equation}
    \begin{split}
        {\color{green}\small{\text{Green lines}}}=\mathcal{L}^{\dagger }{}_{e\uparrow }\mathcal{L}_{g\uparrow }+\mathcal{L}^{\dagger }{}_{f\downarrow }\mathcal{L}_{g\uparrow }+\mathcal{L}^{\dagger }{}_{e\downarrow }\mathcal{L}_{g\downarrow }\\+\mathcal{L}^{\dagger }{}_{f\uparrow }\mathcal{L}_{g\downarrow }+\mathcal{L}^{\dagger }{}_{g\uparrow }\mathcal{L}_{e\uparrow }+\mathcal{L}^{\dagger }{}_{g\uparrow }\mathcal{L}_{f\downarrow }\\+\mathcal{L}^{\dagger }{}_{g\downarrow }\mathcal{L}_{f\uparrow }+\mathcal{L}^{\dagger }{}_{g\downarrow }\mathcal{L}_{e\downarrow }\\
        {\color{blue}\small{\text{Blue lines}}}=-2\mathcal{L}^{\dagger }{}_{e\uparrow }\mathcal{L}_{e\uparrow }-\mathcal{L}^{\dagger }{}_{f\uparrow }\mathcal{L}_{e\uparrow }-\mathcal{L}^{\dagger }{}_{f\downarrow }\mathcal{L}_{e\uparrow }\\-2\mathcal{L}^{\dagger }{}_{e\downarrow }\mathcal{L}_{e\downarrow }-\mathcal{L}^{\dagger }{}_{f\uparrow }\mathcal{L}_{e\downarrow }-\mathcal{L}^{\dagger }{}_{f\downarrow }\mathcal{L}_{e\downarrow }\\-\mathcal{L}^{\dagger }{}_{e\uparrow }\mathcal{L}_{f\uparrow }-\mathcal{L}^{\dagger }{}_{e\downarrow }\mathcal{L}_{f\uparrow }-2\mathcal{L}^{\dagger }{}_{f\downarrow }\mathcal{L}_{f\uparrow }\\-\mathcal{L}^{\dagger }{}_{e\uparrow }\mathcal{L}_{f\downarrow }-\mathcal{L}^{\dagger }{}_{e\downarrow }\mathcal{L}_{f\downarrow }-2\mathcal{L}^{\dagger }{}_{f\uparrow }\mathcal{L}_{f\downarrow }
    \end{split}
\end{equation}
\tikzset{every picture/.style={line width=0.75pt}} 

\begin{tikzpicture}[x=0.75pt,y=0.75pt,yscale=-1.6,xscale=1.6]

\draw    (269.8,122.17) -- (330.18,122.04) ;
\draw    (269.8,142.04) -- (330.18,142.04) ;
\draw [color={rgb, 255:red, 208; green, 49; blue, 2 }  ,draw opacity=1 ] [dash pattern={on 0.84pt off 2.51pt}]  (290.43,122.04) -- (289.93,142.04) ;
\draw [shift={(290.14,133.44)}, rotate = 271.42] [fill={rgb, 255:red, 208; green, 49; blue, 2 }  ,fill opacity=1 ][line width=0.08]  [draw opacity=0] (5.36,-2.57) -- (0,0) -- (5.36,2.57) -- (3.56,0) -- cycle    ;
\draw [color={rgb, 255:red, 214; green, 65; blue, 12 }  ,draw opacity=1 ] [dash pattern={on 0.84pt off 2.51pt}]  (309.81,122.67) -- (309.56,142.92) ;
\draw [shift={(309.72,129.89)}, rotate = 90.7] [fill={rgb, 255:red, 214; green, 65; blue, 12 }  ,fill opacity=1 ][line width=0.08]  [draw opacity=0] (5.36,-2.57) -- (0,0) -- (5.36,2.57) -- (3.56,0) -- cycle    ;
\draw    (379.3,122.17) -- (439.8,122.04) ;
\draw    (379.3,142.04) -- (439.8,142.04) ;
\draw [color={rgb, 255:red, 7; green, 31; blue, 170 }  ,draw opacity=1 ]   (399.97,122.04) -- (399.84,142.54) ;
\draw [shift={(399.89,133.69)}, rotate = 270.37] [fill={rgb, 255:red, 7; green, 31; blue, 170 }  ,fill opacity=1 ][line width=0.08]  [draw opacity=0] (5.36,-2.57) -- (0,0) -- (5.36,2.57) -- (3.56,0) -- cycle    ;
\draw [color={rgb, 255:red, 11; green, 29; blue, 175 }  ,draw opacity=1 ]   (419.39,122.67) -- (419.71,142.79) ;
\draw [shift={(419.5,129.83)}, rotate = 89.08] [fill={rgb, 255:red, 11; green, 29; blue, 175 }  ,fill opacity=1 ][line width=0.08]  [draw opacity=0] (5.36,-2.57) -- (0,0) -- (5.36,2.57) -- (3.56,0) -- cycle    ;
\draw [color={rgb, 255:red, 12; green, 2; blue, 208 }  ,draw opacity=1 ]   (314.71,122.17) .. controls (353.49,101.73) and (360.62,102.73) .. (394.49,122.36) ;
\draw [shift={(356.44,107.24)}, rotate = 180] [fill={rgb, 255:red, 12; green, 2; blue, 208 }  ,fill opacity=1 ][line width=0.08]  [draw opacity=0] (5.36,-2.57) -- (0,0) -- (5.36,2.57) -- (3.56,0) -- cycle    ;
\draw [color={rgb, 255:red, 19; green, 2; blue, 208 }  ,draw opacity=1 ]   (313.43,121.29) .. controls (342.05,91.29) and (374.8,96.04) .. (396.93,122.04) ;
\draw [shift={(352.79,100.7)}, rotate = 0] [fill={rgb, 255:red, 19; green, 2; blue, 208 }  ,fill opacity=1 ][line width=0.08]  [draw opacity=0] (5.36,-2.57) -- (0,0) -- (5.36,2.57) -- (3.56,0) -- cycle    ;
\draw [color={rgb, 255:red, 7; green, 16; blue, 172 }  ,draw opacity=1 ]   (318.04,142.12) .. controls (353.41,153.38) and (364.25,150.33) .. (385.72,142.33) ;
\draw [shift={(349.34,149.24)}, rotate = 2.31] [fill={rgb, 255:red, 7; green, 16; blue, 172 }  ,fill opacity=1 ][line width=0.08]  [draw opacity=0] (5.36,-2.57) -- (0,0) -- (5.36,2.57) -- (3.56,0) -- cycle    ;
\draw    (309.77,181.98) -- (349.9,182.23) ;
\draw    (360.15,181.86) -- (400.27,182.11) ;
\draw [color={rgb, 255:red, 7; green, 29; blue, 199 }  ,draw opacity=1 ]   (269.8,142.04) .. controls (241.62,118.54) and (297.37,127.29) .. (269.74,141.42) ;
\draw [shift={(271.63,128.06)}, rotate = 184.34] [fill={rgb, 255:red, 7; green, 29; blue, 199 }  ,fill opacity=1 ][line width=0.08]  [draw opacity=0] (7.14,-3.43) -- (0,0) -- (7.14,3.43) -- (4.74,0) -- cycle    ;
\draw [color={rgb, 255:red, 208; green, 38; blue, 2 }  ,draw opacity=1 ] [dash pattern={on 0.84pt off 2.51pt}]  (437.52,142.42) .. controls (409.34,118.92) and (465.09,127.67) .. (437.46,141.79) ;
\draw [shift={(439.35,128.44)}, rotate = 184.34] [fill={rgb, 255:red, 208; green, 38; blue, 2 }  ,fill opacity=1 ][line width=0.08]  [draw opacity=0] (7.14,-3.43) -- (0,0) -- (7.14,3.43) -- (4.74,0) -- cycle    ;
\draw [color={rgb, 255:red, 19; green, 2; blue, 208 }  ,draw opacity=1 ]   (304.55,122.42) .. controls (276.37,98.92) and (332.12,107.67) .. (304.49,121.79) ;
\draw [shift={(306.38,108.44)}, rotate = 184.34] [fill={rgb, 255:red, 19; green, 2; blue, 208 }  ,fill opacity=1 ][line width=0.08]  [draw opacity=0] (7.14,-3.43) -- (0,0) -- (7.14,3.43) -- (4.74,0) -- cycle    ;
\draw [color={rgb, 255:red, 208; green, 45; blue, 2 }  ,draw opacity=1 ] [dash pattern={on 0.84pt off 2.51pt}]  (400.03,122.67) .. controls (371.84,99.17) and (427.59,107.92) .. (399.97,122.04) ;
\draw [shift={(401.86,108.69)}, rotate = 184.34] [fill={rgb, 255:red, 208; green, 45; blue, 2 }  ,fill opacity=1 ][line width=0.08]  [draw opacity=0] (7.14,-3.43) -- (0,0) -- (7.14,3.43) -- (4.74,0) -- cycle    ;
\draw [color={rgb, 255:red, 208; green, 45; blue, 2 }  ,draw opacity=1 ] [dash pattern={on 0.84pt off 2.51pt}]  (315.72,182.37) .. controls (342.69,204.69) and (285.41,200.33) .. (315.65,181.74) ;
\draw [shift={(311.53,196.84)}, rotate = 4.99] [fill={rgb, 255:red, 208; green, 45; blue, 2 }  ,fill opacity=1 ][line width=0.08]  [draw opacity=0] (7.14,-3.43) -- (0,0) -- (7.14,3.43) -- (4.74,0) -- cycle    ;
\draw [color={rgb, 255:red, 208; green, 45; blue, 2 }  ,draw opacity=1 ] [dash pattern={on 0.84pt off 2.51pt}]  (396.44,182.91) .. controls (423.41,205.24) and (366.14,200.88) .. (396.38,182.29) ;
\draw [shift={(392.26,197.39)}, rotate = 4.99] [fill={rgb, 255:red, 208; green, 45; blue, 2 }  ,fill opacity=1 ][line width=0.08]  [draw opacity=0] (7.14,-3.43) -- (0,0) -- (7.14,3.43) -- (4.74,0) -- cycle    ;
\draw [color={rgb, 255:red, 208; green, 38; blue, 2 }  ,draw opacity=1 ] [dash pattern={on 0.84pt off 2.51pt}]  (333.78,182.51) .. controls (344.32,195.97) and (367.96,194.51) .. (376.49,183.08) ;
\draw [shift={(356.63,192.11)}, rotate = 180] [fill={rgb, 255:red, 208; green, 38; blue, 2 }  ,fill opacity=1 ][line width=0.08]  [draw opacity=0] (5.36,-2.57) -- (0,0) -- (5.36,2.57) -- (3.56,0) -- cycle    ;
\draw [color={rgb, 255:red, 208; green, 38; blue, 2 }  ,draw opacity=1 ] [dash pattern={on 0.84pt off 2.51pt}]  (331.96,182.88) .. controls (339.6,200.69) and (368.32,204.33) .. (378.32,182.33) ;
\draw [shift={(352.57,197.41)}, rotate = 0] [fill={rgb, 255:red, 208; green, 38; blue, 2 }  ,fill opacity=1 ][line width=0.08]  [draw opacity=0] (5.36,-2.57) -- (0,0) -- (5.36,2.57) -- (3.56,0) -- cycle    ;
\draw [color={rgb, 255:red, 10; green, 20; blue, 173 }  ,draw opacity=1 ]   (325.3,121.64) .. controls (340.99,102.38) and (386.46,124.17) .. (384.56,142.59) ;
\draw [shift={(361.53,117.99)}, rotate = 199.21] [fill={rgb, 255:red, 10; green, 20; blue, 173 }  ,fill opacity=1 ][line width=0.08]  [draw opacity=0] (5.36,-2.57) -- (0,0) -- (5.36,2.57) -- (3.56,0) -- cycle    ;
\draw [color={rgb, 255:red, 11; green, 21; blue, 180 }  ,draw opacity=1 ]   (318.04,142.12) .. controls (341.09,125.22) and (375.93,121.54) .. (380.14,122.38) ;
\draw [shift={(349.42,127.4)}, rotate = 162.33] [fill={rgb, 255:red, 11; green, 21; blue, 180 }  ,fill opacity=1 ][line width=0.08]  [draw opacity=0] (5.36,-2.57) -- (0,0) -- (5.36,2.57) -- (3.56,0) -- cycle    ;
\draw [color={rgb, 255:red, 11; green, 21; blue, 180 }  ,draw opacity=1 ]   (327.04,142.62) .. controls (330,136.5) and (378,124.75) .. (389.14,122.88) ;
\draw [shift={(354.61,131.27)}, rotate = 344.4] [fill={rgb, 255:red, 11; green, 21; blue, 180 }  ,fill opacity=1 ][line width=0.08]  [draw opacity=0] (5.36,-2.57) -- (0,0) -- (5.36,2.57) -- (3.56,0) -- cycle    ;
\draw [color={rgb, 255:red, 7; green, 16; blue, 172 }  ,draw opacity=1 ]   (328.69,142.25) .. controls (348.44,147.25) and (361.19,146.75) .. (379.3,142.04) ;
\draw [shift={(355.56,145.77)}, rotate = 180.86] [fill={rgb, 255:red, 7; green, 16; blue, 172 }  ,fill opacity=1 ][line width=0.08]  [draw opacity=0] (5.36,-2.57) -- (0,0) -- (5.36,2.57) -- (3.56,0) -- cycle    ;
\draw [color={rgb, 255:red, 10; green, 20; blue, 173 }  ,draw opacity=1 ]   (330.18,122.04) .. controls (352.43,109.79) and (373.43,130.7) .. (379.3,142.04) ;
\draw [shift={(359.82,122.6)}, rotate = 204.13] [fill={rgb, 255:red, 10; green, 20; blue, 173 }  ,fill opacity=1 ][line width=0.08]  [draw opacity=0] (5.36,-2.57) -- (0,0) -- (5.36,2.57) -- (3.56,0) -- cycle    ;
\draw [color={rgb, 255:red, 65; green, 117; blue, 5 }  ,draw opacity=1 ]   (295.88,122.5) -- (324.06,181.88) ;
\draw [shift={(308.73,149.57)}, rotate = 64.6] [fill={rgb, 255:red, 65; green, 117; blue, 5 }  ,fill opacity=1 ][line width=0.08]  [draw opacity=0] (5.36,-2.57) -- (0,0) -- (5.36,2.57) -- (3.56,0) -- cycle    ;
\draw [color={rgb, 255:red, 65; green, 117; blue, 5 }  ,draw opacity=1 ]   (318.04,142.12) -- (335.94,181.75) ;
\draw [shift={(325.79,159.29)}, rotate = 65.7] [fill={rgb, 255:red, 65; green, 117; blue, 5 }  ,fill opacity=1 ][line width=0.08]  [draw opacity=0] (5.36,-2.57) -- (0,0) -- (5.36,2.57) -- (3.56,0) -- cycle    ;
\draw [color={rgb, 255:red, 65; green, 117; blue, 5 }  ,draw opacity=1 ]   (409.55,122.11) -- (391.19,181.5) ;
\draw [shift={(401.23,149.03)}, rotate = 107.18] [fill={rgb, 255:red, 65; green, 117; blue, 5 }  ,fill opacity=1 ][line width=0.08]  [draw opacity=0] (5.36,-2.57) -- (0,0) -- (5.36,2.57) -- (3.56,0) -- cycle    ;
\draw [color={rgb, 255:red, 65; green, 117; blue, 5 }  ,draw opacity=1 ]   (393.89,143.19) -- (374.81,181.13) ;
\draw [shift={(385.66,159.57)}, rotate = 116.7] [fill={rgb, 255:red, 65; green, 117; blue, 5 }  ,fill opacity=1 ][line width=0.08]  [draw opacity=0] (5.36,-2.57) -- (0,0) -- (5.36,2.57) -- (3.56,0) -- cycle    ;
\draw [color={rgb, 255:red, 65; green, 117; blue, 5 }  ,draw opacity=1 ]   (390.25,142.34) -- (370.06,182.13) ;
\draw [shift={(379.52,163.48)}, rotate = 296.9] [fill={rgb, 255:red, 65; green, 117; blue, 5 }  ,fill opacity=1 ][line width=0.08]  [draw opacity=0] (5.36,-2.57) -- (0,0) -- (5.36,2.57) -- (3.56,0) -- cycle    ;
\draw [color={rgb, 255:red, 65; green, 117; blue, 5 }  ,draw opacity=1 ]   (291.88,122.5) -- (320.06,181.88) ;
\draw [shift={(306.57,153.45)}, rotate = 244.6] [fill={rgb, 255:red, 65; green, 117; blue, 5 }  ,fill opacity=1 ][line width=0.08]  [draw opacity=0] (5.36,-2.57) -- (0,0) -- (5.36,2.57) -- (3.56,0) -- cycle    ;
\draw [color={rgb, 255:red, 65; green, 117; blue, 5 }  ,draw opacity=1 ]   (313.56,142.88) -- (331.31,181.63) ;
\draw [shift={(323.02,163.52)}, rotate = 245.39] [fill={rgb, 255:red, 65; green, 117; blue, 5 }  ,fill opacity=1 ][line width=0.08]  [draw opacity=0] (5.36,-2.57) -- (0,0) -- (5.36,2.57) -- (3.56,0) -- cycle    ;
\draw [color={rgb, 255:red, 65; green, 117; blue, 5 }  ,draw opacity=1 ]   (414.75,122.89) -- (396.38,182.29) ;
\draw [shift={(405.15,153.93)}, rotate = 287.18] [fill={rgb, 255:red, 65; green, 117; blue, 5 }  ,fill opacity=1 ][line width=0.08]  [draw opacity=0] (5.36,-2.57) -- (0,0) -- (5.36,2.57) -- (3.56,0) -- cycle    ;
\draw [color={rgb, 255:red, 214; green, 65; blue, 12 }  ,draw opacity=1 ] [dash pattern={on 0.84pt off 2.51pt}]  (318.56,143.42) -- (362.69,182.13) ;
\draw [shift={(338.44,160.86)}, rotate = 41.25] [fill={rgb, 255:red, 214; green, 65; blue, 12 }  ,fill opacity=1 ][line width=0.08]  [draw opacity=0] (5.36,-2.57) -- (0,0) -- (5.36,2.57) -- (3.56,0) -- cycle    ;
\draw [color={rgb, 255:red, 214; green, 65; blue, 12 }  ,draw opacity=1 ] [dash pattern={on 0.84pt off 2.51pt}]  (324.04,142.12) -- (368.17,180.82) ;
\draw [shift={(347.16,162.39)}, rotate = 221.25] [fill={rgb, 255:red, 214; green, 65; blue, 12 }  ,fill opacity=1 ][line width=0.08]  [draw opacity=0] (5.36,-2.57) -- (0,0) -- (5.36,2.57) -- (3.56,0) -- cycle    ;
\draw [color={rgb, 255:red, 214; green, 65; blue, 12 }  ,draw opacity=1 ] [dash pattern={on 0.84pt off 2.51pt}]  (379.3,142.04) -- (339.56,182.19) ;
\draw [shift={(358.45,163.11)}, rotate = 314.71] [fill={rgb, 255:red, 214; green, 65; blue, 12 }  ,fill opacity=1 ][line width=0.08]  [draw opacity=0] (5.36,-2.57) -- (0,0) -- (5.36,2.57) -- (3.56,0) -- cycle    ;
\draw [color={rgb, 255:red, 214; green, 65; blue, 12 }  ,draw opacity=1 ] [dash pattern={on 0.84pt off 2.51pt}]  (384.56,142.59) -- (346.56,182.44) ;
\draw [shift={(367.57,160.42)}, rotate = 133.64] [fill={rgb, 255:red, 214; green, 65; blue, 12 }  ,fill opacity=1 ][line width=0.08]  [draw opacity=0] (5.36,-2.57) -- (0,0) -- (5.36,2.57) -- (3.56,0) -- cycle    ;
\draw [color={rgb, 255:red, 208; green, 38; blue, 2 }  ,draw opacity=1 ] [dash pattern={on 0.84pt off 2.51pt}]  (320.96,121.88) .. controls (349.31,114.69) and (383.31,172.69) .. (387.06,182.69) ;
\draw [shift={(359.4,141.46)}, rotate = 47.88] [fill={rgb, 255:red, 208; green, 38; blue, 2 }  ,fill opacity=1 ][line width=0.08]  [draw opacity=0] (5.36,-2.57) -- (0,0) -- (5.36,2.57) -- (3.56,0) -- cycle    ;
\draw [color={rgb, 255:red, 208; green, 38; blue, 2 }  ,draw opacity=1 ] [dash pattern={on 0.84pt off 2.51pt}]  (319.31,122.19) .. controls (347.92,123.75) and (378.56,172.94) .. (382.31,182.94) ;
\draw [shift={(357.99,146.94)}, rotate = 227.95] [fill={rgb, 255:red, 208; green, 38; blue, 2 }  ,fill opacity=1 ][line width=0.08]  [draw opacity=0] (5.36,-2.57) -- (0,0) -- (5.36,2.57) -- (3.56,0) -- cycle    ;
\draw [color={rgb, 255:red, 208; green, 38; blue, 2 }  ,draw opacity=1 ] [dash pattern={on 0.84pt off 2.51pt}]  (380.14,122.38) .. controls (341.56,118.19) and (324.31,170.94) .. (328.06,180.94) ;
\draw [shift={(341.28,141.83)}, rotate = 307.18] [fill={rgb, 255:red, 208; green, 38; blue, 2 }  ,fill opacity=1 ][line width=0.08]  [draw opacity=0] (5.36,-2.57) -- (0,0) -- (5.36,2.57) -- (3.56,0) -- cycle    ;
\draw [color={rgb, 255:red, 208; green, 38; blue, 2 }  ,draw opacity=1 ] [dash pattern={on 0.84pt off 2.51pt}]  (383.39,123.07) .. controls (346.81,125.69) and (327.56,171.63) .. (331.31,181.63) ;
\draw [shift={(348.84,141.2)}, rotate = 131.57] [fill={rgb, 255:red, 208; green, 38; blue, 2 }  ,fill opacity=1 ][line width=0.08]  [draw opacity=0] (5.36,-2.57) -- (0,0) -- (5.36,2.57) -- (3.56,0) -- cycle    ;

\draw (420.65,146.85) node [anchor=north west][inner sep=0.75pt]  [font=\tiny]  {$f,\sigma =\frac{1}{2}$};
\draw (415.15,106.32) node [anchor=north west][inner sep=0.75pt]  [font=\tiny]  {$f,\sigma =-\frac{1}{2}$};
\draw (243.05,101.44) node [anchor=north west][inner sep=0.75pt]  [font=\tiny]  {$e,\sigma =\frac{1}{2}$};
\draw (239.22,143.44) node [anchor=north west][inner sep=0.75pt]  [font=\tiny]  {$e,\sigma =-\frac{1}{2}$};
\draw (327.04,197.64) node [anchor=north west][inner sep=0.75pt]  [font=\tiny]  {$g,\sigma =\pm \frac{1}{2}$};
\draw (310,214.69) node [anchor=north west][inner sep=0.75pt]   [align=left] {\tiny{\textbf{Scatterings in I method}}};

\end{tikzpicture}

This choice of interactions also yields us the only hybridization terms we considered in our model.  
\begin{equation}
    \begin{split}
    \label{eq:our_model}
        d^{\dagger }{}_\uparrow c_{\frac{1}{2}\uparrow}+d^{\dagger }{}_\downarrow c_{\frac{1}{2}\downarrow}-d^{\dagger }{}_\downarrow c_{-\frac{1}{2}\uparrow}-d^{\dagger }{}_\uparrow c_{-\frac{1}{2}\downarrow}\\+c^{\dagger }{}_{\frac{1}{2}\uparrow} d_\uparrow+c^{\dagger }{}_{\frac{1}{2}\downarrow} d_\downarrow-c^{\dagger }{}_{-\frac{1}{2}\uparrow} d_\downarrow-c^{\dagger }{}_{-\frac{1}{2}\downarrow} d_\uparrow
    \end{split}
\end{equation}
\begin{equation}
    \begin{split}
       {\color{green}\small{\text{Green lines}}}= \mathcal{L}^{\dagger }{}_{e\uparrow} \mathcal{L}_{g\downarrow}+\mathcal{L}^{\dagger }{}_{g\downarrow} \mathcal{L}_{e\uparrow}+\mathcal{L}^{\dagger }{}_{f\downarrow} \mathcal{L}_{g\downarrow}\\+\mathcal{L}^{\dagger }{}_{g\downarrow} \mathcal{L}_{f\downarrow}+\mathcal{L}^{\dagger }{}_{e\uparrow} \mathcal{L}_{g\uparrow}+\mathcal{L}^{\dagger }{}_{g\uparrow} \mathcal{L}_{e\uparrow}\\+\mathcal{L}^{\dagger }{}_{f\downarrow} \mathcal{L}_{g\uparrow}+\mathcal{L}^{\dagger }{}_{g\uparrow} \mathcal{L}_{f\downarrow}\\
       {\color{blue}\small{\text{Blue lines}}}=-\mathcal{L}^{\dagger }{}_{e\uparrow} \mathcal{L}_{e\downarrow}-\mathcal{L}^{\dagger }{}_{e\downarrow} \mathcal{L}_{e\uparrow}-\mathcal{L}^{\dagger }{}_{f\downarrow} \mathcal{L}_{e\downarrow}\\-\mathcal{L}^{\dagger }{}_{e\downarrow} \mathcal{L}_{f\downarrow}-2 \mathcal{L}^{\dagger }{}_{f\downarrow} \mathcal{L}_{e\uparrow}-2 \mathcal{L}^{\dagger }{}_{e\uparrow} \mathcal{L}_{f\downarrow}\\-\mathcal{L}^{\dagger }{}_{f\uparrow} \mathcal{L}_{e\uparrow}-\mathcal{L}^{\dagger }{}_{e\uparrow} \mathcal{L}_{f \uparrow}-2 \mathcal{L}^{\dagger }{}_{e\uparrow} \mathcal{L}_{e\uparrow}\\-\mathcal{L}^{\dagger }{}_{f\uparrow} \mathcal{L}_{f\downarrow}-\mathcal{L}^{\dagger }{}_{f\downarrow} \mathcal{L}_{f\uparrow}-2 \mathcal{L}^{\dagger }{}_{f\downarrow} \mathcal{L}_{f\downarrow}
    \end{split}
\end{equation}
The above Lindbladian construction may be seen graphically as the following, 
\tikzset{every picture/.style={line width=0.75pt}} 

\begin{tikzpicture}[x=0.75pt,y=0.75pt,yscale=-1.55,xscale=1.55]

\draw    (249.8,102.17) -- (310.18,102.04) ;
\draw    (249.8,122.04) -- (310.18,122.04) ;
\draw [color={rgb, 255:red, 74; green, 96; blue, 226 }  ,draw opacity=1 ]   (270.43,102.04) -- (269.93,122.04) ;
\draw [shift={(270.14,113.44)}, rotate = 271.42] [fill={rgb, 255:red, 74; green, 96; blue, 226 }  ,fill opacity=1 ][line width=0.08]  [draw opacity=0] (5.36,-2.57) -- (0,0) -- (5.36,2.57) -- (3.56,0) -- cycle    ;
\draw [color={rgb, 255:red, 74; green, 91; blue, 226 }  ,draw opacity=1 ]   (289.81,102.67) -- (289.56,122.92) ;
\draw [shift={(289.72,109.89)}, rotate = 90.7] [fill={rgb, 255:red, 74; green, 91; blue, 226 }  ,fill opacity=1 ][line width=0.08]  [draw opacity=0] (5.36,-2.57) -- (0,0) -- (5.36,2.57) -- (3.56,0) -- cycle    ;
\draw    (359.3,102.17) -- (419.8,102.04) ;
\draw    (359.3,122.04) -- (419.8,122.04) ;
\draw [color={rgb, 255:red, 74; green, 96; blue, 226 }  ,draw opacity=1 ]   (379.97,102.04) -- (379.84,122.54) ;
\draw [shift={(379.89,113.69)}, rotate = 270.37] [fill={rgb, 255:red, 74; green, 96; blue, 226 }  ,fill opacity=1 ][line width=0.08]  [draw opacity=0] (5.36,-2.57) -- (0,0) -- (5.36,2.57) -- (3.56,0) -- cycle    ;
\draw [color={rgb, 255:red, 74; green, 91; blue, 226 }  ,draw opacity=1 ]   (399.39,102.67) -- (399.71,122.79) ;
\draw [shift={(399.5,109.83)}, rotate = 89.08] [fill={rgb, 255:red, 74; green, 91; blue, 226 }  ,fill opacity=1 ][line width=0.08]  [draw opacity=0] (5.36,-2.57) -- (0,0) -- (5.36,2.57) -- (3.56,0) -- cycle    ;
\draw [color={rgb, 255:red, 208; green, 38; blue, 2 }  ,draw opacity=1 ] [dash pattern={on 0.84pt off 2.51pt}]  (294.71,102.17) .. controls (333.49,81.73) and (340.62,82.73) .. (374.49,102.36) ;
\draw [shift={(336.44,87.24)}, rotate = 180] [fill={rgb, 255:red, 208; green, 38; blue, 2 }  ,fill opacity=1 ][line width=0.08]  [draw opacity=0] (5.36,-2.57) -- (0,0) -- (5.36,2.57) -- (3.56,0) -- cycle    ;
\draw [color={rgb, 255:red, 208; green, 45; blue, 2 }  ,draw opacity=1 ] [dash pattern={on 0.84pt off 2.51pt}]  (293.43,101.29) .. controls (322.05,71.29) and (354.8,76.04) .. (376.93,102.04) ;
\draw [shift={(332.79,80.7)}, rotate = 0] [fill={rgb, 255:red, 208; green, 45; blue, 2 }  ,fill opacity=1 ][line width=0.08]  [draw opacity=0] (5.36,-2.57) -- (0,0) -- (5.36,2.57) -- (3.56,0) -- cycle    ;
\draw [color={rgb, 255:red, 74; green, 83; blue, 226 }  ,draw opacity=1 ]   (304.72,122.17) .. controls (325.41,127.59) and (343.3,126.43) .. (359.72,122.64) ;
\draw [shift={(333.82,125.86)}, rotate = 181.09] [fill={rgb, 255:red, 74; green, 83; blue, 226 }  ,fill opacity=1 ][line width=0.08]  [draw opacity=0] (5.36,-2.57) -- (0,0) -- (5.36,2.57) -- (3.56,0) -- cycle    ;
\draw [color={rgb, 255:red, 74; green, 83; blue, 226 }  ,draw opacity=1 ]   (298.04,122.12) .. controls (333.41,133.38) and (344.25,130.33) .. (365.72,122.33) ;
\draw [shift={(329.34,129.24)}, rotate = 2.31] [fill={rgb, 255:red, 74; green, 83; blue, 226 }  ,fill opacity=1 ][line width=0.08]  [draw opacity=0] (5.36,-2.57) -- (0,0) -- (5.36,2.57) -- (3.56,0) -- cycle    ;
\draw    (289.77,161.98) -- (329.9,162.23) ;
\draw    (340.15,161.86) -- (380.27,162.11) ;
\draw [color={rgb, 255:red, 65; green, 117; blue, 5 }  ,draw opacity=1 ]   (284.99,121.92) -- (303.41,161.6) ;
\draw [shift={(294.79,143.03)}, rotate = 245.1] [fill={rgb, 255:red, 65; green, 117; blue, 5 }  ,fill opacity=1 ][line width=0.08]  [draw opacity=0] (5.36,-2.57) -- (0,0) -- (5.36,2.57) -- (3.56,0) -- cycle    ;
\draw [color={rgb, 255:red, 65; green, 117; blue, 5 }  ,draw opacity=1 ]   (289.99,122.17) -- (309.84,162.11) ;
\draw [shift={(298.62,139.54)}, rotate = 63.58] [fill={rgb, 255:red, 65; green, 117; blue, 5 }  ,fill opacity=1 ][line width=0.08]  [draw opacity=0] (5.36,-2.57) -- (0,0) -- (5.36,2.57) -- (3.56,0) -- cycle    ;
\draw [color={rgb, 255:red, 5; green, 117; blue, 6 }  ,draw opacity=1 ]   (374.87,122.51) -- (359.71,161.42) ;
\draw [shift={(366.78,143.27)}, rotate = 291.29] [fill={rgb, 255:red, 5; green, 117; blue, 6 }  ,fill opacity=1 ][line width=0.08]  [draw opacity=0] (5.36,-2.57) -- (0,0) -- (5.36,2.57) -- (3.56,0) -- cycle    ;
\draw [color={rgb, 255:red, 28; green, 117; blue, 5 }  ,draw opacity=1 ]   (379.84,122.54) -- (365.9,162.29) ;
\draw [shift={(373.83,139.68)}, rotate = 109.32] [fill={rgb, 255:red, 28; green, 117; blue, 5 }  ,fill opacity=1 ][line width=0.08]  [draw opacity=0] (5.36,-2.57) -- (0,0) -- (5.36,2.57) -- (3.56,0) -- cycle    ;
\draw [color={rgb, 255:red, 28; green, 117; blue, 5 }  ,draw opacity=1 ]   (297.87,122.67) -- (340.87,161.67) ;
\draw [shift={(320.4,143.11)}, rotate = 222.21] [fill={rgb, 255:red, 28; green, 117; blue, 5 }  ,fill opacity=1 ][line width=0.08]  [draw opacity=0] (5.36,-2.57) -- (0,0) -- (5.36,2.57) -- (3.56,0) -- cycle    ;
\draw [color={rgb, 255:red, 32; green, 117; blue, 5 }  ,draw opacity=1 ]   (303.99,122.67) -- (347.74,161.67) ;
\draw [shift={(323.7,140.24)}, rotate = 41.71] [fill={rgb, 255:red, 32; green, 117; blue, 5 }  ,fill opacity=1 ][line width=0.08]  [draw opacity=0] (5.36,-2.57) -- (0,0) -- (5.36,2.57) -- (3.56,0) -- cycle    ;
\draw [color={rgb, 255:red, 74; green, 93; blue, 226 }  ,draw opacity=1 ]   (249.8,122.04) .. controls (221.62,98.54) and (277.37,107.29) .. (249.74,121.42) ;
\draw [shift={(251.63,108.06)}, rotate = 184.34] [fill={rgb, 255:red, 74; green, 93; blue, 226 }  ,fill opacity=1 ][line width=0.08]  [draw opacity=0] (7.14,-3.43) -- (0,0) -- (7.14,3.43) -- (4.74,0) -- cycle    ;
\draw [color={rgb, 255:red, 74; green, 93; blue, 226 }  ,draw opacity=1 ]   (417.52,122.42) .. controls (389.34,98.92) and (445.09,107.67) .. (417.46,121.79) ;
\draw [shift={(419.35,108.44)}, rotate = 184.34] [fill={rgb, 255:red, 74; green, 93; blue, 226 }  ,fill opacity=1 ][line width=0.08]  [draw opacity=0] (7.14,-3.43) -- (0,0) -- (7.14,3.43) -- (4.74,0) -- cycle    ;
\draw [color={rgb, 255:red, 28; green, 117; blue, 5 }  ,draw opacity=1 ]   (360.18,122.04) -- (322.24,162.17) ;
\draw [shift={(340.25,143.12)}, rotate = 313.39] [fill={rgb, 255:red, 28; green, 117; blue, 5 }  ,fill opacity=1 ][line width=0.08]  [draw opacity=0] (5.36,-2.57) -- (0,0) -- (5.36,2.57) -- (3.56,0) -- cycle    ;
\draw [color={rgb, 255:red, 28; green, 117; blue, 5 }  ,draw opacity=1 ]   (365.74,122.92) -- (328.9,162.23) ;
\draw [shift={(349.3,140.46)}, rotate = 133.14] [fill={rgb, 255:red, 28; green, 117; blue, 5 }  ,fill opacity=1 ][line width=0.08]  [draw opacity=0] (5.36,-2.57) -- (0,0) -- (5.36,2.57) -- (3.56,0) -- cycle    ;
\draw [color={rgb, 255:red, 208; green, 45; blue, 2 }  ,draw opacity=1 ] [dash pattern={on 0.84pt off 2.51pt}]  (284.55,102.42) .. controls (256.37,78.92) and (312.12,87.67) .. (284.49,101.79) ;
\draw [shift={(286.38,88.44)}, rotate = 184.34] [fill={rgb, 255:red, 208; green, 45; blue, 2 }  ,fill opacity=1 ][line width=0.08]  [draw opacity=0] (7.14,-3.43) -- (0,0) -- (7.14,3.43) -- (4.74,0) -- cycle    ;
\draw [color={rgb, 255:red, 208; green, 45; blue, 2 }  ,draw opacity=1 ] [dash pattern={on 0.84pt off 2.51pt}]  (380.03,102.67) .. controls (351.84,79.17) and (407.59,87.92) .. (379.97,102.04) ;
\draw [shift={(381.86,88.69)}, rotate = 184.34] [fill={rgb, 255:red, 208; green, 45; blue, 2 }  ,fill opacity=1 ][line width=0.08]  [draw opacity=0] (7.14,-3.43) -- (0,0) -- (7.14,3.43) -- (4.74,0) -- cycle    ;
\draw [color={rgb, 255:red, 208; green, 45; blue, 2 }  ,draw opacity=1 ] [dash pattern={on 0.84pt off 2.51pt}]  (295.72,162.37) .. controls (322.69,184.69) and (265.41,180.33) .. (295.65,161.74) ;
\draw [shift={(291.53,176.84)}, rotate = 4.99] [fill={rgb, 255:red, 208; green, 45; blue, 2 }  ,fill opacity=1 ][line width=0.08]  [draw opacity=0] (7.14,-3.43) -- (0,0) -- (7.14,3.43) -- (4.74,0) -- cycle    ;
\draw [color={rgb, 255:red, 208; green, 45; blue, 2 }  ,draw opacity=1 ] [dash pattern={on 0.84pt off 2.51pt}]  (376.44,162.91) .. controls (403.41,185.24) and (346.14,180.88) .. (376.38,162.29) ;
\draw [shift={(372.26,177.39)}, rotate = 4.99] [fill={rgb, 255:red, 208; green, 45; blue, 2 }  ,fill opacity=1 ][line width=0.08]  [draw opacity=0] (7.14,-3.43) -- (0,0) -- (7.14,3.43) -- (4.74,0) -- cycle    ;
\draw [color={rgb, 255:red, 208; green, 38; blue, 2 }  ,draw opacity=1 ] [dash pattern={on 0.84pt off 2.51pt}]  (313.78,162.51) .. controls (324.32,175.97) and (347.96,174.51) .. (356.49,163.08) ;
\draw [shift={(336.63,172.11)}, rotate = 180] [fill={rgb, 255:red, 208; green, 38; blue, 2 }  ,fill opacity=1 ][line width=0.08]  [draw opacity=0] (5.36,-2.57) -- (0,0) -- (5.36,2.57) -- (3.56,0) -- cycle    ;
\draw [color={rgb, 255:red, 208; green, 38; blue, 2 }  ,draw opacity=1 ] [dash pattern={on 0.84pt off 2.51pt}]  (311.96,162.88) .. controls (319.6,180.69) and (348.32,184.33) .. (358.32,162.33) ;
\draw [shift={(332.57,177.41)}, rotate = 0] [fill={rgb, 255:red, 208; green, 38; blue, 2 }  ,fill opacity=1 ][line width=0.08]  [draw opacity=0] (5.36,-2.57) -- (0,0) -- (5.36,2.57) -- (3.56,0) -- cycle    ;
\draw [color={rgb, 255:red, 74; green, 83; blue, 226 }  ,draw opacity=1 ]   (308.88,101.96) .. controls (335.97,89.17) and (357.65,111.73) .. (360.18,122.04) ;
\draw [shift={(336.35,99.97)}, rotate = 22.85] [fill={rgb, 255:red, 74; green, 83; blue, 226 }  ,fill opacity=1 ][line width=0.08]  [draw opacity=0] (5.36,-2.57) -- (0,0) -- (5.36,2.57) -- (3.56,0) -- cycle    ;
\draw [color={rgb, 255:red, 74; green, 83; blue, 226 }  ,draw opacity=1 ]   (305.3,101.64) .. controls (320.99,82.38) and (366.46,104.17) .. (364.56,122.59) ;
\draw [shift={(341.53,97.99)}, rotate = 199.21] [fill={rgb, 255:red, 74; green, 83; blue, 226 }  ,fill opacity=1 ][line width=0.08]  [draw opacity=0] (5.36,-2.57) -- (0,0) -- (5.36,2.57) -- (3.56,0) -- cycle    ;
\draw [color={rgb, 255:red, 74; green, 83; blue, 226 }  ,draw opacity=1 ]   (298.04,122.12) .. controls (321.09,105.22) and (355.93,101.54) .. (360.14,102.38) ;
\draw [shift={(329.42,107.4)}, rotate = 162.33] [fill={rgb, 255:red, 74; green, 83; blue, 226 }  ,fill opacity=1 ][line width=0.08]  [draw opacity=0] (5.36,-2.57) -- (0,0) -- (5.36,2.57) -- (3.56,0) -- cycle    ;
\draw [color={rgb, 255:red, 74; green, 83; blue, 226 }  ,draw opacity=1 ]   (303.99,122.67) .. controls (331.93,107.96) and (363.51,102.17) .. (366.1,102.93) ;
\draw [shift={(331.67,110.94)}, rotate = 342.02] [fill={rgb, 255:red, 74; green, 83; blue, 226 }  ,fill opacity=1 ][line width=0.08]  [draw opacity=0] (5.36,-2.57) -- (0,0) -- (5.36,2.57) -- (3.56,0) -- cycle    ;
\draw [color={rgb, 255:red, 226; green, 39; blue, 13 }  ,draw opacity=1 ] [dash pattern={on 0.84pt off 2.51pt}]  (267.77,101.65) -- (289.77,161.98) ;
\draw [shift={(279.25,133.13)}, rotate = 249.97] [fill={rgb, 255:red, 226; green, 39; blue, 13 }  ,fill opacity=1 ][line width=0.08]  [draw opacity=0] (5.36,-2.57) -- (0,0) -- (5.36,2.57) -- (3.56,0) -- cycle    ;
\draw [color={rgb, 255:red, 226; green, 39; blue, 13 }  ,draw opacity=1 ] [dash pattern={on 0.84pt off 2.51pt}]  (273.65,101.41) -- (295.65,161.74) ;
\draw [shift={(283.66,128.85)}, rotate = 69.97] [fill={rgb, 255:red, 226; green, 39; blue, 13 }  ,fill opacity=1 ][line width=0.08]  [draw opacity=0] (5.36,-2.57) -- (0,0) -- (5.36,2.57) -- (3.56,0) -- cycle    ;
\draw [color={rgb, 255:red, 226; green, 39; blue, 13 }  ,draw opacity=1 ] [dash pattern={on 0.84pt off 2.51pt}]  (310.18,102.04) -- (348.79,161.58) ;
\draw [shift={(330.25,132.99)}, rotate = 237.04] [fill={rgb, 255:red, 226; green, 39; blue, 13 }  ,fill opacity=1 ][line width=0.08]  [draw opacity=0] (5.36,-2.57) -- (0,0) -- (5.36,2.57) -- (3.56,0) -- cycle    ;
\draw [color={rgb, 255:red, 226; green, 39; blue, 13 }  ,draw opacity=1 ] [dash pattern={on 0.84pt off 2.51pt}]  (305.3,101.64) -- (343.91,161.18) ;
\draw [shift={(323.03,128.98)}, rotate = 57.04] [fill={rgb, 255:red, 226; green, 39; blue, 13 }  ,fill opacity=1 ][line width=0.08]  [draw opacity=0] (5.36,-2.57) -- (0,0) -- (5.36,2.57) -- (3.56,0) -- cycle    ;
\draw [color={rgb, 255:red, 226; green, 39; blue, 13 }  ,draw opacity=1 ] [dash pattern={on 0.84pt off 2.51pt}]  (400.66,103.19) -- (374.96,162.42) ;
\draw [shift={(387.25,134.09)}, rotate = 293.46] [fill={rgb, 255:red, 226; green, 39; blue, 13 }  ,fill opacity=1 ][line width=0.08]  [draw opacity=0] (5.36,-2.57) -- (0,0) -- (5.36,2.57) -- (3.56,0) -- cycle    ;
\draw [color={rgb, 255:red, 226; green, 39; blue, 13 }  ,draw opacity=1 ] [dash pattern={on 0.84pt off 2.51pt}]  (395.66,101.52) -- (369.96,160.75) ;
\draw [shift={(383.96,128.48)}, rotate = 113.46] [fill={rgb, 255:red, 226; green, 39; blue, 13 }  ,fill opacity=1 ][line width=0.08]  [draw opacity=0] (5.36,-2.57) -- (0,0) -- (5.36,2.57) -- (3.56,0) -- cycle    ;
\draw [color={rgb, 255:red, 226; green, 39; blue, 13 }  ,draw opacity=1 ] [dash pattern={on 0.84pt off 2.51pt}]  (373.44,102.44) -- (313.29,163.58) ;
\draw [shift={(342.38,134.01)}, rotate = 314.53] [fill={rgb, 255:red, 226; green, 39; blue, 13 }  ,fill opacity=1 ][line width=0.08]  [draw opacity=0] (5.36,-2.57) -- (0,0) -- (5.36,2.57) -- (3.56,0) -- cycle    ;
\draw [color={rgb, 255:red, 226; green, 39; blue, 13 }  ,draw opacity=1 ] [dash pattern={on 0.84pt off 2.51pt}]  (369.44,101.61) -- (309.29,162.75) ;
\draw [shift={(341.4,130.11)}, rotate = 134.53] [fill={rgb, 255:red, 226; green, 39; blue, 13 }  ,fill opacity=1 ][line width=0.08]  [draw opacity=0] (5.36,-2.57) -- (0,0) -- (5.36,2.57) -- (3.56,0) -- cycle    ;

\draw (400.65,126.85) node [anchor=north west][inner sep=0.75pt]  [font=\tiny]  {$f,\sigma =\frac{1}{2}$};
\draw (395.15,86.32) node [anchor=north west][inner sep=0.75pt]  [font=\tiny]  {$f,\sigma =-\frac{1}{2}$};
\draw (223.05,89.44) node [anchor=north west][inner sep=0.75pt]  [font=\tiny]  {$e,\sigma =\frac{1}{2}$};
\draw (219.22,127.44) node [anchor=north west][inner sep=0.75pt]  [font=\tiny]  {$e,\sigma =-\frac{1}{2}$};
\draw (307.04,177.64) node [anchor=north west][inner sep=0.75pt]  [font=\tiny]  {$g,\sigma =\pm \frac{1}{2}$};
\draw (290,190.69) node [anchor=north west][inner sep=0.75pt]   [align=left] {\tiny{\textbf{Scatterings in II method}}};

\end{tikzpicture}
This above preserve the total number in Kondo-relevant angular momentum channels since we find the coefficients by calculating the commutators as $[({\color{green}\small{\text{Green lines}}}+{\color{blue}\small{\text{Blue lines}}})_{no-flip},(n_{j_m\sigma}+n_{d\sigma})]=0$ and flip terms are odd under parity and the coefficients evaluated as $\lbrace ({\color{green}\small{\text{Green lines}}}+{\color{blue}\small{\text{Blue lines}}})_{flip},(n_{j_m\sigma}+n_{d\sigma}) \rbrace=0$,All blue lines correspond to the negative sign scattering terms and green correspond to positive scattering terms. Red dots are the terms left out because we preserve number in $j_m=\pm \frac{1}{2}$. In dot, there is no dissipation; hence, we did not consider the onsite dissipation for the dot operators and no flip terms, which will introduce the non-Hermitian spin-orbit interaction in dot, which is out of the scope of the current work. These terms may arise in controlled dissipation, and these signs by this way of calculation also tell us about even and odd under parity,which also ensure commutation with the metric operator.  
\begin{equation}
    \begin{split}
    \label{eq:II_method}
       {\color{green}\small{\text{Green lines}}}+{\color{blue}\small{\text{Blue lines}}}=\\-d^{\dagger }{}_\uparrow c_{-\frac{1}{2}\downarrow}+d^{\dagger }{}_\uparrow c_{-\frac{1}{2}\downarrow}+d^{\dagger }{}_\uparrow c_{\frac{1}{2}\uparrow}-d^{\dagger }{}_\downarrow c_{\frac{1}{2}\uparrow}\\-c^{\dagger }{}_{-\frac{1}{2}\downarrow} d_1+c^{\dagger }{}_{-\frac{1}{2}\downarrow} d_\downarrow+c^{\dagger }{}_{\frac{1}{2}\uparrow} d_\uparrow-c^{\dagger }{}_{\frac{1}{2}\uparrow} d_\downarrow
    \end{split}
\end{equation}
 This hybridization can be written as the following by multiplying a prefactor as $(iV+\omega)$ then we get the following terms. This is a laborious calculation of commutators and anticommutators but in a straightforward way. The prefactor adds to the original SIAM in $(j_m,\sigma)$ basis will give the following terms. We also bring back the k-index for bath operators and coefficients. 
\begin{equation}
    \begin{split}
    \label{eq:hyb1}
        H^{hyb}=\sum_{k}V_{k}(c^\dagger_{k\frac{1}{2}\uparrow}d_{\uparrow}+c^\dagger_{k,-\frac{1}{2}\downarrow}d_{\downarrow}+hc)\\
        \frac{i}{2}\mathcal{L}^\dagger_{\sigma}\mathcal{L}_{\sigma'}=\sum_{k}(-\frac{V_{k}}{2}+\frac{i}{2}\omega_k) \bigg(c^{\dagger }{}_{k\frac{1}{2}\uparrow} d_\uparrow+c^{\dagger }{}_{k,-\frac{1}{2}\downarrow} d_\downarrow\\-c^{\dagger }{}_{k\frac{1}{2}\uparrow} d_\downarrow-c^{\dagger }{}_{k,-\frac{1}{2}\downarrow} d_\uparrow+h.c \bigg)\\
        \implies H^{hyb}+ \frac{i}{2}\mathcal{L}^\dagger_{\sigma}\mathcal{L}_{\sigma'}=H^{\rm hyb}_{\rm flip}+H^{\rm hyb}_{\rm nflip}
    \end{split}
\end{equation}
\emph{Out of the two possibilities \ref{eq:our_model} and \ref{eq:II_method} we chose the first one from Lindbladians to map to the type of Anderson model\cite{loure} however, we write both these following hybridization terms, which are $\mathcal{PT}$-symmetric and both yield the same eigenvalues.}  \\
\textbf{I Method}
\begin{equation}
    \begin{split}
    \label{eq:hyb2}
         H^{hyb}_{nflip} &=
     \sum_{k}X_k\left(c_{k,+\frac{1}{2}\uparrow}^\dagger d_\uparrow + c_{k,\frac{1}{2}\downarrow}^\dagger d_\downarrow + {\rm h.c}\right)  \\
    H^{hyb}_{flip} &=
      -\sum_{k}  X_k^*\left(c_{k,-\frac{1}{2}\uparrow}^\dagger d_\downarrow + c_{k,-\frac{1}{2}\downarrow}^\dagger d_\uparrow + {\rm h.c}\right)
    \end{split}
\end{equation}
\textbf{II method}
\begin{equation}
    \begin{split}
    \label{eq:hyb1}
         H^{hyb}_{nflip} &=
     \sum_{k}X_k\left(c_{k,+\frac{1}{2}\uparrow}^\dagger d_\uparrow + c_{k,-\frac{1}{2}\downarrow}^\dagger d_\downarrow + {\rm h.c}\right)  \\
    H^{hyb}_{flip} &=
      \sum_{k}  X_k^*\left(c_{k,\frac{1}{2}\uparrow}^\dagger d_\downarrow + c_{k,-\frac{1}{2}\downarrow}^\dagger d_\uparrow + {\rm h.c}\right)
    \end{split}
\end{equation}
So, we can see \textbf{I method} shown here below analyses to compare with the \cite{loure} work, \textbf{II method} is indeed consistent with the \cite{zarea}. However, the additional terms brought through Lndbladian are to explore the interplay between the dissipation and parity-breaking interaction. Only in bath this parity breaking can  be verified with commutation of only bath $S_z$ operator  $[(n_{\pm\frac{1}{2}\uparrow}-n_{\pm\frac{1}{2}\downarrow}),H^{hyb}_{nflip}+H^{hyb}_{flip}]\ne 0$.This condition is satisfied by both hybridization terms, and they yield the same eigenvalues since there is no angular momentum-dependent coefficient for hybridization. However, from \textbf{I method} it is convenient to find simpler metric operators(no-phase attached to d operators in eigenvector and so on..) to map to the non-Hermitian studied version of the model to compare the results. \\
Here we show from the following gauge choices the \ref{eq:hyb2} and \ref{eq:hyb1} are equivalent,
\begin{equation}
    \begin{split}
       \text{for no-flip}: c_{k\frac{1}{2}\uparrow}=e^{i\theta}c_{kL\uparrow}\\
        c_{k-\frac{1}{2}\downarrow}=e^{-i\theta}c_{kL\downarrow}\\
        \text{for flip}: c_{k\frac{1}{2}\uparrow}=-e^{-i\theta}c_{kR\uparrow}\\
        c_{k-\frac{1}{2}\downarrow}=-e^{i\theta}c_{kR\downarrow}
    \end{split}
\end{equation}
from this choice we get the following unitary and show both of the hybridization yield the same model\ref{eq:Canderson} which we used for various calculations. 
\begin{equation}
    \begin{split}
        \mathcal{U}=\frac{1}{\sqrt{2}}\begin{pmatrix}
        e^{-i\theta}&e^{i\theta}\\
        e^{-i\theta} & -e^{i\theta}
        \end{pmatrix}\\
        \mathcal{U}\begin{pmatrix}
        d_{\uparrow}\\d_{\downarrow}
        \end{pmatrix}=\begin{pmatrix}
        d_{+}\\d_{-}
        \end{pmatrix}
    \end{split}
\end{equation}
Thus, the full non-Hermitian
Hamiltonian in the angular momentum basis then becomes
\begin{equation}
    H^{\mbox{\tiny{NH}}}_{\mbox{\tiny{SIAM}}} = H_0+H^{\rm hyb}_{\rm flip}+H^{\rm hyb}_{\rm nflip}+H_{\rm RSO}+H_d\,.
\end{equation}
Equation~\ref{eq:hsrom} is off-diagonal in the 'm' basis, but a simple transformation to the total angular momentum basis, namely $j_m=m+\sigma$ gives the following:
\begin{equation}
\label{eq:hsrojmA}
    H_{RSO}=\lambda\sum_{kj_m} \left(c^\dag_{kj_m\uparrow} 
    c^{\phantom{\dag}}_{kj_m\downarrow} + {\rm h.c.}\right)
\end{equation}
which is diagonal in the $j_m$ basis. The kinetic energy term and the  hybridization terms may also be rewritten in the same way, and we get the following:
\begin{align}
    H_{0}&=
        \sum_{kj_m\sigma}\tilde{\epsilon}_{k}c^{\dagger}_{kj_m\sigma}c_{kj_m\sigma}  \label{eq:H0jmA} \\
     H^{hyb}_{nflip} &=
     \sum_{k}X_k\left(c_{k,+\frac{1}{2}\uparrow}^\dagger d_\uparrow + c_{k,\frac{1}{2}\downarrow}^\dagger d_\downarrow + {\rm h.c}\right)  \\
    H^{hyb}_{flip} &=
      -\sum_{k}  X_k^*\left(c_{k,-\frac{1}{2}\uparrow}^\dagger d_\downarrow + c_{k,-\frac{1}{2}\downarrow}^\dagger d_\uparrow + {\rm h.c}\right) \label{eq:flipnfA}
\end{align}
where $X_k=(V_k+i\omega_k)/2\sqrt{k}$. 
Note that the Lindbladian formalism allows us to choose coefficients of the hybridization in a specific way, that maintains the ${\mathcal{PT}}$-symmetry of the Hamiltonian.
The RSOC term (equation~\ref{eq:hsrojmA}) is  off-diagonal in the spin-index, so we can combine it with equation~\ref{eq:H0jmA} through a unitary rotation
of the $\sigma_z$ basis into a `chiral' basis, namely,
\begin{equation}
    c_{kj_mh}=\frac{1}{\sqrt{2}}(c_{kj_m\uparrow}+hc_{kj_m\downarrow})
\end{equation}
where $h=\pm$, to a form that is diagonal in the chiral quantum number ($h$).
So, we get
\begin{equation}
    H_0+H_{\rm RSO}= \sum_{kj_mh}\tilde{\epsilon}_{kh}
    c^\dag_{kj_mh}c^{\phantom{\dag}}_{kj_mh}
\end{equation}
where $\tilde{\epsilon}_{kh}=\tilde{\epsilon}_{k}+h\lambda$.
In this rotated basis ($j_m\sigma\longrightarrow j_mh$),
and with the identification of $j_{m}=\pm 1/2$ as the left (L) and right (R) channels respectively, the model may be interpreted as a system with one interacting quantum dot hybridizing with two conduction electron baths. The full Hamiltonian written below reflects such an interpretation.
\begin{equation}
   \begin{split}
    &H =\sum_{kh\eta}\tilde{\epsilon}_{kh}^{\phantom{\dagger}} c_{kh\eta}^\dagger c_{kh\eta}^{\phantom{\dagger}}  \\
 & +\sum_k X_k\left[ \left(c^{\dagger}_{kL+}+ c^{\dagger}_{kL-}\right)d_{\uparrow}^{\phantom{\dagger}}+ \left(c^{\dagger}_{kL+} - c^{\dagger}_{kL-}\right)d_{\downarrow}^{\phantom{\dagger}}+ {\rm h.c.}\right]\\
 & -\sum_k X_k^*\left[ \left(c^{\dagger}_{kR+}+ c^{\dagger}_{kR-}\right)d_{\downarrow}^{\phantom{\dagger}}+ \left(c^{\dagger}_{kR+} - c^{\dagger}_{kR-}\right)d_{\uparrow}^{\phantom{\dagger}}+ {\rm h.c.}\right]\\
 &+\sum_{\sigma}\epsilon_d n_{d\sigma}
    + Un_{d\uparrow}n_{d\downarrow}
    \end{split}
    \label{eq:SOandersonA}
\end{equation}
where $\eta=L,R$ is the channel index. 
Defining a rotation of the spin basis on the dot $\left(d_+ \;\;d_-\right)^T = U\left(d_\uparrow \;\;d_\downarrow\right)^T$, where 
the unitary rotation is given by $U=(\sigma_z + \sigma_x)/\sqrt{2}$, the model Hamiltonian may be condensed
into a form which appears very similar to a conventional Anderson impurity
model connected to two baths, namely:
\begin{equation}
   \begin{split}
    H &=\sum_{kh\eta}\tilde{\epsilon}_{kh}^{\phantom{\dagger}} c_{kh\eta}^\dagger c_{kh\eta}^{\phantom{\dagger}}  
  +\sum_{k\eta h} X_{k\eta h}\left( c^{\dagger}_{k\eta h}d_h^{\phantom{\dagger}}+ {\rm h.c.}\right)\\
 &+\epsilon_d \sum_{h} n_{dh}
    + Un_{d+}n_{d-}
    \end{split}
    \label{eq:CandersonA}
\end{equation}
where $X_{kLh} = \sqrt{2}|X_k|e^{i\phi_k}$ and $X_{kRh} = -hX_{kLh}^*$.\\


\bibliography{ref}
\end{document}